\documentclass[final]{pasj01}

\Received{$\langle$reception date$\rangle$}

\Accepted{$\langle$acception date$\rangle$}

\Published{$\langle$publication date$\rangle$}

\usepackage{graphicx}
\usepackage{rotating}
\usepackage{afterpage}
\usepackage{txfonts}
\usepackage[sort&compress]{natbib}
\bibpunct{(}{)}{;}{a}{}{,}
\newcommand{\um}{$\mu$m}

\newcommand{\kms}{km\thinspace s$^{-1}$}

\def\arcsec{\hbox{$^{\prime\prime}$}}
\def\utw{\smash{\rlap{\lower5pt\hbox{$\sim$}}}}
\def\udtw{\smash{\rlap{\lower6pt\hbox{$\approx$}}}}

\def\farcs{\hbox{$.\!\!^{\prime\prime}$}}

\def\Msun{\hbox{\it M$_\odot$}}

\def\Mbol{\hbox{\it M$_{bol}$}}

\newcommand{\Ks}{{\it K$_{\rm s}$}}

\newcommand{\Aks}{{\it A$_{\it K_{\rm s}}$}}

\def\simgr{\mathrel{\hbox{\rlap{\hbox{\lower4pt\hbox{$\sim$}}}\hbox{$>$}}}}

\def\vlsr{\hbox{V$_{\rm LSR}$}}
\def\Vlsr{\hbox{V$_{\rm LSR}$}}

\def\vexp{\hbox{V$_{\rm exp}$}}

\begin{document}

\title{86 GHz SiO masers in Galactic Centre OH/IR stars.}
\author{Messineo, M.$^{1,3}$,  
Sjouwerman, L. O.$^2$,
Habing, H. J.$^3$,
and Omont, A.$^4$
}

\altaffiltext{1}{
Key Laboratory for Researches in Galaxies and Cosmology, University of Science and Technology of China, 
Chinese Academy of Sciences, Hefei, Anhui, 230026, China}
\altaffiltext{2}{
National Radio Astronomy Observatory, PO Box O, Socorro NM 87801, USA
}
\altaffiltext{3}{
Leiden Observatory, PO Box 9513, 2300 RA Leiden, The Netherlands
}
\altaffiltext{4}{
Institut d'Astrophysique de Paris, CNRS, 98bis boulevard Arago, 75014 Paris, France
}


\maketitle

\begin{abstract}
We present  results on a search for 86.243 GHz SiO 
($J = 2 \rightarrow 1, v = 1$) 
maser emission toward 67 OH/IR stars located near the Galactic Centre. 
We detected  32 spectral peaks, of which 28 correspond to SiO maser 
lines arising from the envelopes of these OH/IR stars. 
 In OH/IR stars, we obtained  an SiO maser detection rate of about 40\%.
 We  serendipitously detected  two other lines  from OH/IR stars  
at $\approx$ 86.18 GHz,  which could be due to a CCS-molecule transition at 86.181 GHz 
or probably to an highly excited OH molecular transition at 86.178 GHz.
The detection rate of 86 GHz maser emission 
is found to be about 60\% for sources with 
The Midcourse Space Experiment (MSX) $A-E < 2.5$ mag; 
but it drops to 25\% for the reddest OH/IR stars with MSX $A-E > 2.5$
mag.
This supports the  hypothesis by \citet{messineo02} that  the SiO masers 
are primarily found in relatively thinner circumstellar material.
\end{abstract}

\section{\label{introduction}Introduction}
At the end of their life, Asymptotic Giant Branch  (AGB) stars
($\approx1 < M_*< \approx 8 \, M_\odot$)  enter a phase of intense mass loss at rates 
of typically 10$^{-7}$ to 10$^{-4}$ M$_\odot$ yr$^{-1}$ \citep{herwig05}.
A variety of names are used in literature to
indicate  thermal pulsing AGB stars, either referring to their
pulsation properties (e.g., semiregular, Miras, long-period variables),
or to their envelope properties (optically thick or thin envelopes),
or masers.
Indeed, their circumstellar envelopes  can exhibit maser
emission  \citep[e.g., from SiO,  H$_2$O,  and OH molecules,][]{habing96}. 
SiO maser emission originates close to the stellar photosphere, 
inside the dust formation zone. 
H$_2$O maser spots form further out, in the acceleration region,
and 1612 MHz OH masers originate in the cooler part of the envelope
where the expanding shell has reached  a  terminal velocity (known as V$_{\rm exp}$).
\citet{vanderveen88} and \citet{lewis89}
analyzed the colors of IRAS sources, and suggested an evolutionary sequence
of increasing mass-loss rate and maser occurrence (from SiO, via
H$_2$O, to OH masers) with redder colors.
More recently, instead of a sequence of mass-loss rate, 
\citet{sjouwerman09} interprets this sequence to be more  likely
a shell thickness or shell density estimator. 
Regardless, the observed sequence indicates an empirical 
stellar type; it goes from shell-less evolved late-type stars,
Mira-type stars with thin shells and silicate features in
emission, through optically obscured thick-shell OH/IR stars
with silicate features in absorption, and ends in planetary
nebulae.
SiO maser transitions are easily detected  at 43 
($J = 1 \rightarrow 0, v = 1$ and $v = 2$) GHz and at 86 
GHz ($J = 2 \rightarrow 1, v = 1$)  toward
oxygen-rich AGBs, typically in objects with the 
silicate feature in emission.
The relative strength of the SiO maser lines varies
with source type \citep[Mira versus OH/IR][]{lane82,nyman86,nyman93},
suggesting that the SiO maser pumping mechanism strongly depends
on mass-loss rate (shell thickness) and luminosity; individually, line ratios 
also vary with stellar pulsation phase \citep{stroh18}.
Note that the classic term 'Mira' indicates optically 
visible long-period variables ($> \approx 100$ d)
with large amplitudes, while, today, the term 'OH/IR star'
is often used to indicate   optically thick long period variables (LPVs)
with mass-loss rateabove $10^{-5}$ \Msun\ yr$^{-1}$,
independently of masers \citep[e.g.,][]{blommaert18,habing96}
even though they were discovered and 
classified because  of OH maser emission.

Here, we present spectra of 86 GHz SiO maser lines of a 
sample of 67  OH/IR stars  with OH maser emission. The data set was obtained using the IRAM 30m telescope 
and the data analysis was carried out in the year 2002.  
Since the strength and occurrence of masers vary in phase  with the 
infrared stellar pulsation cycle of these long-period variables, it is valuable to perform and keep track of repeated observations. 
 The main  purpose of this communication is to provide an   
 historical record for these detections, 
 as the raw data are not stored in any archive.
 This dataset was taken to  verify the
  hypothesis made by \citet{messineo02} on their color selection of
 targets for the 86 GHz SiO maser search.
 The authors selected Mira-like stars with bluer color to avoid genuine
 OH/IR stars, and any known OH emitter was removed from their list. 
 This avoidance was based on the idea that SiO maser emission
 was more likely to occur (and perhaps more
 strongly) in relatively thin-shell Miras than in  
 thick-shell OH/IR stars. 
 Evidence was based, though, on  few observations 
 \citep[][and references therein]{messineo02}.

 Along with the SiO masers towards OH/IR stars, a few other 
 serendipitously detected lines are reported.
In Sect.\ \ref{targets} we describe the sample and  observations.
In Sect.\ \ref{analysis}, the 86 GHz detections are presented and commented on where appropriate.

\section{The sample and 86 GHz observations}
\label{targets} 
\label {data}	
The list of 67 AGB OH/IR stars observed is presented in Tables 
\ref{table:detections} (detections) and \ref{table:non-detections} (non-detections).
This sample of sources comes  from the catalogues of OH/IR stars near the 
Galactic Centre (GC) from \citet[comprising 134 sources,][]{lindqvist92}, 
with additional sources from \citet{sjouwerman98}. These OH/IR stars are among the  
brightest  and reddest infrared sources and had  already been detected with the IRAS 
mid-infrared satellite \citep[][]{habing96}. The targeted OH/IR stars were selected 
by having ISOGAL measurements at 7 and 15 \um\ \citep{ortiz02}, i.e., they are located 
in fields observed with ISOCAM on board  the ESA/ISO satellite 
\citep{omont03,schuller03}. 

At the time of these observations,
the sample of OH/IR stars by Lindqvist comprised 134 stars concentrated
in the central one degree. 
The sample of Bulge+Disk OH/IR stars from the blind survey
of \citet{sevenster97} and \citet{sevenster01} comprised  only about 
700 sources\footnote{ Nowadays, \citet{engels15} report about 
2341  stars with OH masers. There is an ongoing new survey of the Galactic plane,
and \citet{qiao18} reported on 161 new OH masers in  the 
central 5 degrees}.
Furthermore, the Lindqvist sample had been reidentified  with 
ISOGAL data. Therefore, due to the stellar density distribution 
and available data, it was straightforward 
to compare our comprehensive Mira-like sample  towards the inner Galaxy, 
with  a sample of OH/IR stars on the Galactic Centre.

The observations were carried out with the IRAM 30-m telescope (Pico Veleta,
Spain) between December 2001 and April 2002  (15h awarded to IRAM program 144-01, 
P.I. Messineo).   
The observing technique and setup as well as the data reduction and analysis 
were identical to those described by \citet{messineo02}.
The IRAM pointing coordinates were based on  The Deep Near Infrared Survey of the southern sky  (DENIS) 
 and typically accurate within 1\arcsec\ \citep{epchtein94}.
IRAM telescope pointing errors are typically within 2-4\arcsec, while the  
86 GHz beam full  at width half maximum (FWHM) is 29\arcsec.   
Two orthogonal linear polarized receivers were tuned  to simultaneously 
observe the SiO maser line at a rest frequency of 
86.24335 GHz  ($J = 2 \rightarrow 1, v = 1$). The receiver output was
combined to obtain total intensity spectra. 
For each receiver we used in parallel the low resolution filter bank 
(3.5 \kms\ spectral resolution and 890 \kms\ total velocity coverage)
and the autocorrelator (1.1 \kms\  of resolution and   973 \kms\ of coverage).
Observations were made in wobbler switching mode with a
wobbler throw of 100-200\arcsec. The individual on-source integration
time was set  between 12 and 24 minutes per source, depending on the system
temperature (110-250 K). The conversion factor from antenna temperature 
to flux density is 6.2 Jy K$^{-1}$.

\begin{sidewaystable*}
\vspace*{+0.4cm} 
\caption{\label{table:detections}  Compilation of parameters and aliases of OH/IR stars detected at 86 GHz.}
\begin{center}

{\tiny
\begin{tabular}{@{\extracolsep{-.09in}}lccrrrrrll|lrr|rr|llll}
\hline
\hline
   &              &                &             &           &   &                      &                      &                 &                & \multicolumn{3}{c}{\citet{engels15}}   & &  &   &  & \\
ID &RA$^X$        &DEC$^X$         &V$_{\rm LSR}$&T$_{\rm a}$&rms&  A                   &FWHM                  & Obs. Date       & {Alias1$^+$}   & Alias2  &\Vlsr  & \vexp & Per & $\Delta$ K & References & Note\\
   &[J2000 ]      &[J2000]         &[\kms]       & [K] &[K]    & [K\ \kms]          &[\kms]                & [yymmdd]            &                                   &  &  [\kms]& [\kms]& [days] & [mag]          & \\
\hline
 1$^a$ & 17:44:28.391 & $-$29:26:33.76 &  {\slshape (414.31)}&  0.104&  0.014&   0.54  $\pm$   0.08 &  11.15   $\pm$  2.41 & 011214/020326     &          LI004&OH359.437$-$0.051&  $-$138.20&    15.30&    $..$&  $..$&                1,3,4 &   "SiO"(?) velocity differs from OH\\
     2 & 17:44:34.987 & $-$29:04:35.47 &     $-$6.08&  0.133&  0.015&   0.51  $\pm$   0.03 &   3.46   $\pm$  0.27 &          020326       &          LI033&OH359.762+00.12&    $-$5.55&    14.73&    $..$&  $..$&                  1,5  \\
     3 & 17:44:39.721 & $-$29:16:45.88 &    $-$73.43&  0.085&  0.021&   0.23  $\pm$   0.06 &   2.43   $\pm$  0.73 &          011219       &          LI014&OH359.598+0.000&   $-$73.75&    19.05&   664 &   2.30&              1,3,4,5  \\
     4 & 17:44:46.003 & $-$29:06:13.61 &    $-$58.23&  0.047&  0.013&   0.17  $\pm$   0.05 &   3.75   $\pm$  1.20 &          020319       &          LI032&OH359.760+0.072&   $-$54.90&    20.60&   676 &   2.09&                1,3,4  \\
     5 & 17:44:51.262 & $-$29:09:45.58 &   $-$103.85&  0.089&  0.022&   0.51  $\pm$   0.12 &   6.77   $\pm$  1.99 &          020319       &          LI027&OH359.719+0.025&  $-$102.35&    22.25&   669 &   1.79&                1,3,4  \\
     6 & 17:44:56.881 & $-$29:13:25.36 &   $-$148.39&  0.053&  0.033&   0.49  $\pm$   0.06 &   7.17   $\pm$  1.00 &          020326       &          LI022&OH359.68$-$0.02~~&  $-$149.97&    21.63&    $..$&  $..$&                  1,5  \\
     7 & 17:45:03.241 & $-$29:07:12.69 &    $-$26.72&  0.078&  0.019&   0.35  $\pm$   0.07 &   5.33   $\pm$  1.08 &          020319       &          LI038&OH359.778+0.010&   $-$27.95&    21.05&   572 &   1.19&            1,3,4,6,7  \\
     8 & 17:45:05.333 & $-$29:17:13.31 &   $-$140.78&  0.077&  0.013&   0.45  $\pm$   0.10 &   5.69   $\pm$  1.71 &          011221       &          LI018&OH359.640$-$0.084&  $-$141.65&    14.45&   546 &   0.91&                1,3,4  \\
     9$^{b,*}$ & 17:45:11.549 & $-$28:38:37.36 &  {\slshape (154.69)}&              0.064&  0.013&   0.32  $\pm$   0.09 &   3.60   $\pm$  1.15 &          020319       &          LI101&OH0.200+0.233~~&   $-$67.05&    15.85&   825 &   2.43&              1,3,4,5  & 86.178 GHz OH \vlsr$\sim-65$ \kms\\ 
    10 & 17:45:12.959 & $-$29:12:54.58 &     28.68&  0.098&  0.011&   0.32  $\pm$   0.12 &   3.54   $\pm$  2.35 &          011229       &          LI026&OH359.716$-$0.070&    29.40&    22.25&   691 &   1.94&                1,3,4  \\
    11 & 17:45:13.942 & $-$28:47:43.22 &     21.07&  0.055&  0.009&   0.20  $\pm$   0.03 &   3.36   $\pm$  0.58 &          020328       &          LI086&OH008+0.15~~~~~&    21.83&    21.03&   616 &   2.24&                  1,7  \\
    12 & 17:45:13.988 & $-$29:06:56.30 &    $-$12.60&  0.069&  0.015&   0.23  $\pm$   0.05 &   2.98   $\pm$  0.79 &          020319       &          LI042&OH359.803$-$0.021&   $-$13.45&    22.40&   838 &   0.90&                1,3,4  \\
    13 & 17:45:14.303 & $-$29:07:20.78 &    $-$72.35&  0.027&  0.005&   0.28  $\pm$   0.05 &   7.00   $\pm$  1.28 &          020328       &          Sj004&OH359.797$-$0.025&   $-$71.20&    19.30&   547 &   2.01&              2,5,6,7  \\
    14 & 17:45:17.870 & $-$29:05:53.63 &    $-$53.88&  0.065&  0.023&   0.24  $\pm$   0.04 &   3.70   $\pm$  0.53 &          020326       &          LI047&OH359.83$-$0.02~~&   $-$52.97&    20.63&   660 &   1.56&              1,5,6,7  \\
    15 & 17:45:19.358 & $-$29:14:05.94 &    $-$28.89&  0.058&  0.011&   0.21  $\pm$   0.08 &   3.35   $\pm$  1.45 &          011229       &          LI025&OH359.711$-$0.100&   $-$32.10&    19.45&   686 &   1.38&              1,3,4,5  \\
    16 & 17:45:29.397 & $-$28:39:35.03 &     62.35&  0.036&  0.013&   0.16  $\pm$   0.04 &   4.07   $\pm$  1.25 &          020319       &          LI104&OH0.221+0.168~~&    61.80&    19.20&   697 &   2.48&              1,3,4,5  \\
    17 & 17:45:31.502 & $-$28:46:22.04 &    $-$53.88&  0.053&  0.011&   0.15  $\pm$   0.03 &   2.66   $\pm$  0.67 &          020326       &          LI091&OH0.13+0.10~~~~&   $-$52.37&    11.40&   458 &  $..$&                    1  \\
    18$^c$&{\slshape 17:45:32.089}&{\slshape $-$28:46:19.42}&     43.89&  0.047&  0.011&   0.24  $\pm$   0.05 &   5.75   $\pm$  1.73 &          020326       &     V*V4502Sgr&    $..$&    $..$&    $..$&  $..$&  $..$   &  &       in beam of LI091 (\#17)   \\
    19$^{d}$ & 17:45:34.833 & $-$29:06:02.45 &      3.69&  0.076&  0.020&   0.49  $\pm$   0.08 &   5.21   $\pm$  1.05 &          020319       &          LI050&OH359.855$-$0.078&     4.10&    21.85&   611 &   1.64&              1,3,4,6  \\
    20$^{d,*}$ & 17:45:34.833 & $-$29:06:02.45 &    {\slshape (222.04)}&  0.062&  0.020&   0.45  $\pm$   0.10 &   7.77   $\pm$  2.13 &          020319      &               &&$..$&                  $..$&  $..$ & $..$ &$..$ &86.178 GHz OH \vlsr$\sim+2$ \kms \\
    21 & 17:45:48.497 & $-$29:10:45.01 &    $-$31.07&  0.068&  0.017&   0.20  $\pm$   0.05 &   2.69   $\pm$  0.94 &          020319       &          LI045&OH359.814$-$0.162&   $-$31.70&    20.40&   547 &   1.85&            1,3,4,6,7  \\
    22 & 17:45:56.130 & $-$28:39:27.00 &    102.55&  0.056&  0.013&   0.21  $\pm$   0.04 &   3.56   $\pm$  0.75 &          020319       &          LI109&OH0.274+0.086~~&   103.00&    24.40&   706 &   2.13&                1,3,4  \\
    23 & 17:46:15.001 & $-$28:44:17.31 &     37.37&  0.043&  0.013&   0.16  $\pm$   0.19 &   3.64   $\pm$  5.39 &          020319       &          LI106&OH0.241$-$0.014~~&    35.00&    18.35&   535 &   0.94&                1,3,4  \\
    24 & 17:46:19.597 & $-$29:00:41.72 &      8.04&  0.043&  0.013&   0.19  $\pm$   0.04 &   4.01   $\pm$  0.69 &          020331       &          Sj073&OH0.016$-$0.171~~&     8.20&    19.30&   581 &   2.70&                  2,7  \\
    25 & 17:46:22.161 & $-$28:46:22.51 &   $-$109.28&  0.044&  0.014&   0.15  $\pm$   0.05 &   4.78   $\pm$  1.27 &          020319       &          LI105&OH0.225$-$0.055~~&  $-$106.15&    16.60&   525 &   1.52&              1,3,4,7  \\
    26 & 17:46:28.757 & $-$28:53:19.54 &     41.71&  0.054&  0.014&   0.27  $\pm$   0.12 &   4.69   $\pm$  2.42 &   020319/020328       &          LI094&OH0.138$-$0.136~~&    40.20&    21.10&   810 &   1.40&              1,3,4,5  \\
    27 & 17:46:31.282 & $-$28:39:48.49 &     38.45&  0.095&  0.015&   0.39  $\pm$   0.07 &   3.89   $\pm$  0.81 &          020319       &          LI113&OH0.336$-$0.027~~&    38.00&    17.00&   514 &   1.01&                1,3,4  \\
    28 & 17:46:33.150 & $-$28:45:00.61 &     13.47&  0.110&  0.020&   0.53  $\pm$   0.08 &   4.54   $\pm$  0.71 &          020319       &          LI108&OH0.265$-$0.078~~&    12.15&    17.75&   595 &   1.13&                1,3,4  \\
    29 & 17:46:38.540 & $-$29:08:03.48 &    146.00&  0.049&  0.010&   0.09  $\pm$   0.02 &   1.74   $\pm$  0.55 &          020331       &          Sj041&OH359.947$-$0.294&   143.90&    21.10&    $..$&  $..$&                    2  \\
    30 & 17:47:02.211 & $-$28:45:55.84 &    $-$10.43&  0.084&  0.016&   0.35  $\pm$   0.06 &   4.32   $\pm$  0.89 &          020319       &          LI110&OH0.307$-$0.176~~&    $-$8.50&    20.40&   657 &   1.56&                1,3,4  \\
    31 & 17:47:21.298 & $-$28:39:23.26 &     92.77&  0.038&  0.014&   0.23  $\pm$   0.05 &   5.54   $\pm$  1.23 &          020319       &          LI119&OH0.437$-$0.179~~&    96.20&    17.60&   744 &   2.31&                1,3,4  \\
    32 & 17:47:24.052 & $-$28:32:43.01 &    147.08&  0.060&  0.010&   0.32  $\pm$   0.04 &   4.76   $\pm$  0.68 &   011219/020206       &          LI127&OH0.536$-$0.130~~&   146.20&    23.30&   669 &   1.48&                1,3,4  \\

\hline
\end{tabular}
}
\end{center}
\begin{tabnote}
{\bf References:}  1=\citet{lindqvist92};~ 2=\citet{sjouwerman98};
3=\citet{wood98}; 4=\citet{ortiz02}; 
5=\citet{vanhollebeke06}; 6=\citet{glass01}; 7=\citet{matsunaga09}.\\
{\bf Notes:}  \Vlsr\ and \vexp\ are from the 1612 MHz OH measurements and are the average of the measurements listed in the catalog of \citet{engels15}. \\
\footnotemark[$a$] At the location of OH359.437$-$0.051 (LI004, ID=1) 
we detected a broad high-velocity line
at \Vlsr=414.31 \kms\ (FWHM=11.16 \kms), assuming a rest frequency of the SiO transition.
The actual 1612 MHz OH maser velocity is  \Vlsr=$-138.20$ \kms, and
its corresponding anticipated SiO maser is not detected at 86 GHz around this 1612 MHz OH velocity.
LI004 is the brightest mid-infrared star with   [4.5] = 10.4 mag, [8.0]=6.1 mag, [15.0]=4.24 mag. 
 Only two other  8 \um\  GLIMPSE stars fall within the IRAM telescope beam, 
but they are much fainter  with [8.0]=7.70 and 7.80 mag. 
With a rest frequency of about 86.084 GHz (which is uncertain to a few MHz), the nature of this serendipitous line is undetemined.\\
\footnotemark[$b$] At the position of the OH/IR star  LI101/OH0.200+0.233  we detected a line 
at \Vlsr=154.69 \kms\ (FWHM=3.6 \kms), while the 1612 MHz OH maser line is  at \Vlsr= $-$67.40 \kms. We think that this is a detection of OH instead of SiO; see the note further down.\\
\footnotemark[$c$] We  serendipitously detected another SiO maser  when pointing to OH0.129+0.103 (ID=\#17, LI091).
The OH/IR star has \Vlsr=$-52.75$ \kms\  and coincides with ISOGAL-P J174534.8-290602, 
the brightest 15 \um\ star \citep[3.6 mag,][]{ortiz02}. 
The second maser (\Vlsr=43.89 \kms, ID=\#18) detected in the same beam 
likely belongs to the known long-period variable V* V4502 Sgr (8\farcs8 distant).
Both variables are detected by GLIMPSE with [8.0]=5.47 and 6.65 mag, and 
[3.6-8.0]= 1.66 and 0.63 mag, respectively.\\
\footnotemark[$d$] At the position of OH/IR star LI050   we detected two lines; one at \Vlsr=  
3.69 \kms\ (FWHM=5.21 \kms) and another line at \Vlsr= 222.04 \kms\ (FWHM=7.77 \kms). 
The 3.69 \kms\ line is an  SiO detection of
OH359.855$-$0.078 with a 1612 MHz OH maser velocity detemined at \Vlsr=3.95 \kms\ 
\citep{lindqvist92}. In addition to SiO we think we also detect OH in this object; 
see the next note below.\\
\footnotemark[$*$] The velocities  of two detected  lines (ID=9 and 20) 
differ from the expected LI101 and LI050 1612 MHz OH maser velocities by about 220 \kms\ when assuming 
they originate as the SiO $(v=1, J = 2 \rightarrow 1)$ maser.
Alternatively it  suggests a possible line from the targeted source at a 
frequency of about 86.18 GHz. We propose these detections, instead of
from SiO, originate from an highly excited 
OH $^2\Pi_{3/2}$ J$=\frac{17}{2}$ ($v=1$) F=8$^+$-8$^-$ main line transition at 86.178 GHz. 
There are no other bright mid-infrared stars 
([8.0]$<$ 8.9 mag) at these positions.
\footnotemark[$+$] Alias names are taken from \citet{ortiz02} and SIMBAD.
\footnotemark[$X$] IRAM telescope pointing coordinates.
\end{tabnote}
\end{sidewaystable*}

\begin{sidewaystable*}
\vspace*{+0.2cm} 
\caption{\label{table:non-detections}  Compilation of parameters and aliases of OH/IR stars undetected at 86 GHz.}
\begin{center}
{\tiny
\begin{tabular}{@{\extracolsep{-.08in}}lccrll|lrr|rcllrrr}
\hline
\hline
   &              &                &             &                       &                & \multicolumn{3}{c}{\citet{engels15}}   & &  &   &  & \\
ID &RA$^X$        &DEC$^X$         &{\rm rms}    & {\rm Obs. Date}       & {Alias1$^+$}   & Alias2  &\Vlsr  & \vexp & Per & $\Delta$ K & References \\
   &[J2000 ]      &[J2000]         &      [K]    &  [{\rm yymmdd}]       &                &         & [\kms]& [\kms]& [days] & [mag]          & \\
\hline
    33 & 17:43:45.521 & $-$29:26:16.22       & 0.012 &          020326       &          LI001&OH359.360+00.08&  $-$211.50&    12.33&    $..$&  $..$&                   1  \\
    34 & 17:43:54.019 & $-$29:25:23.41       & 0.012 &          020326       &          LI002&OH359.388+0.066&  $-$129.10&    17.60&    $..$&  $..$&                   1  \\
    35 & 17:44:14.988 & $-$28:45:05.98       & 0.018 &          020326       &          LI076&OH000.000+00.35&   124.03&     9.40&   477 &   1.06&               1,3,5  \\
    36 & 17:44:34.541 & $-$29:10:37.17       & 0.018 &          011221       &          LI021&OH359.68+0.07~~&   $-$24.50&    18.67&   698 &   2.05&             1,3,4,5  \\
    37 & 17:44:44.462 & $-$29:05:38.29       & 0.013 &          020319       &          LI035&OH359.765+0.082&   110.95&    17.60&   552 &   2.22&             1,3,4,5  \\
    38 & 17:44:47.983 & $-$29:06:49.82       & 0.014 &          020319       &          LI031&OH359.755+0.061&    37.90&    18.75&    $..$&  $..$&             1,3,4,5  \\
    39 & 17:44:54.199 & $-$29:13:44.94       & 0.018 &          011221       &          LI020&OH359.669$-$0.019&   $-$83.80&    17.50&   481 &   1.01&             1,3,4,5  \\
    40 & 17:44:57.779 & $-$29:20:42.50       & 0.015 &          011219       &          LI012&OH359.576$-$0.091&   $-$55.50&    18.60&   672 &   2.94&             1,3,4,5  \\
    41 & 17:45:01.740 & $-$29:02:49.99       & 0.022 &          020401       &          Sj011&OH359.838+0.052&   $-$68.20&    12.30&   444 &   0.85&               2,5,6  \\
    42 & 17:45:06.992 & $-$29:03:34.16       & 0.014 &          020319       &          LI048&OH359.837+0.030&   $-$75.35&     7.95&   399 &   1.07&         1,3,4,5,6,7  \\
    43 & 17:45:10.467 & $-$29:18:11.77       & 0.017 &          011219       &          LI017&OH359.636$-$0.108&  $-$138.05&    22.30&   847 &   2.73&               1,3,4  \\
    44 & 17:45:12.480 & $-$28:40:44.11       & 0.014 &          020319       &          LI096&OH0.173+0.211~~&    46.40&    17.05&   514 &   1.57&             1,3,4,5  \\
    45 & 17:45:13.109 & $-$29:09:36.19       & 0.014 &          011214       &          LI034&OH359.763$-$0.042&   120.30&    13.90&   457 &   1.24&             1,3,4,6  \\
    46 & 17:45:13.890 & $-$29:15:28.55       & 0.017 &          020319       &          LI023&OH359.681$-$0.095&   $-$98.85&    19.45&   759 &   2.95&               1,3,4  \\
    47 & 17:45:16.460 & $-$29:15:37.55       & 0.010 &          020206       &          LI024&OH359.684$-$0.104&   $-$59.35&    16.85&   535 &   1.26&               1,3,4  \\
    48 & 17:45:18.119 & $-$29:18:04.03       & 0.017 &          011221       &          LI019&OH359.652$-$0.131&  $-$188.35&    20.35&   671 &   2.03&               1,3,4  \\
    49 & 17:45:29.279 & $-$29:07:04.22       & 0.008 &          020331       &          Sj008&OH359.830$-$0.070&   $-$82.70&    18.20&   567 &   1.36&               2,6,7  \\
    50 & 17:45:29.508 & $-$29:09:16.60       & 0.015 &          020328       &          LI040&OH359.80$-$0.09~~&    $-$3.90&    17.97&   626 &   2.40&               1,5,7  \\
    51 & 17:45:33.331 & $-$29:11:23.17       & 0.029 &          020326       &          LI037&OH359.78$-$0.12~~&    71.80&    13.03&   284 &   1.00&                 1,6  \\
    52 & 17:45:33.449 & $-$28:43:45.08       & 0.012 &          020331       &          Sj001&OH0.170+0.119~~&   116.00&    22.70&   999 &   1.86&                 2,7  \\
    53 & 17:45:34.439 & $-$29:12:54.22       & 0.026 &          020401       &          Sj002&OH359.757$-$0.136&   $-$10.80&    20.30&    $..$&  $..$&                 2,5  \\
    54 & 17:45:46.609 & $-$28:32:40.13       & 0.014 &          020319       &          LI115&OH0.352+0.175~~&    10.80&    18.20&   661 &   2.24&               1,3,4  \\
    55 & 17:45:54.228 & $-$28:31:46.74       & 0.014 &          020319       &          LI116&OH0.379+0.159~~&   139.30&    15.30&   985 &   3.03&               1,3,4  \\
    56 & 17:46:01.113 & $-$29:01:24.20       & 0.019 &          011214       &          LI070&OH359.97$-$0.12~~&    $-$8.37&    19.30&    $..$&  $..$&               1,3,4  \\
    57 & 17:46:11.722 & $-$28:59:32.10       & 0.014 &          020401       &          Sj075&OH0.017$-$0.137~~&   108.30&    18.90&   396 &   1.19&               2,6,7  \\
    58 & 17:46:14.529 & $-$28:36:39.49       & 0.014 &          020319       &          LI114&OH0.349+0.053~~&    31.80&    14.20&   669 &   2.26&               1,3,4  \\
    59 & 17:46:15.407 & $-$28:55:42.71       & 0.075 &          020326       &          LI087&OH0.08$-$0.12~~~~&    50.83&    14.20&    $..$&  $..$&                   1  \\
    60 & 17:46:15.840 & $-$28:56:32.39       & 0.025 &          020331       &          SJ092&OH0.067$-$0.123~~&    35.70&    17.40&   534 &   1.30&               2,6,7  \\
    61 & 17:46:24.790 & $-$29:00:02.38       & 0.014 &          020319       &          LI080&OH0.036$-$0.182~~&   152.80&    18.50&   669 &   2.60&             1,3,4,7  \\
    62 & 17:46:30.737 & $-$28:31:31.87       & 0.010 &          011229       &          LI121&OH0.452+0.046~~&    87.70&    10.20&   339 &   0.94&               1,3,4  \\
    63 & 17:46:31.780 & $-$28:35:40.81       & 0.014 &          020319       &          LI117&OH0.395+0.008~~&   200.10&    13.70&   461 &   1.15&               1,3,4  \\
    64 & 17:46:35.478 & $-$28:58:58.98       & 0.012 &          020328       &          LI085&OH000.071$-$00.20&   112.98&    13.55&    $..$&  $..$&                   1  \\
    65 & 17:46:42.330 & $-$28:33:26.14       & 0.018 &          020319       &          LI120&OH0.447$-$0.006~~&  $-$186.90&    13.10&   445 &   1.85&               1,3,4  \\
    66 & 17:46:44.848 & $-$28:34:59.30       & 0.021 &          020401       &          LI118&OH0.430$-$0.027~~&    31.80&    18.70&    $..$&  $..$&                   1  \\
    67 & 17:46:47.871 & $-$28:47:15.04       & 0.017 &          020319       &          LI107&OH0.261$-$0.143~~&    25.75&     5.55&    $..$&  $..$&               1,3,4  \\
    68 & 17:46:59.057 & $-$28:16:58.51       & 0.027 &          020328       &          LI132&OH0.713+0.084~~&    96.80&    20.50&    $..$&  $..$&                   1  \\
    69 & 17:47:39.822 & $-$28:35:48.69       & 0.013 &          020326       &          LI126&OH0.523$-$0.206~~&    10.80&    21.60&  1050 &   2.14&                 1,3  \\

\hline
\end{tabular}
}

\end{center}
\begin{tabnote}
{\bf References:}  1=\citet{lindqvist92};~ 2=\citet{sjouwerman98};
3=\citet{wood98}; 4=\citet{ortiz02}; 
5=\citet{vanhollebeke06}; 6=\citet{glass01};  7=\citet{matsunaga09}.
{\bf Notes:}  \Vlsr\ and \vexp\ are from the 1612 MHz OH measurements and are the average of the measurements listed in the catalog of \citet{engels15}. \\
\footnotemark[$+$] Alias names are taken from \citet{ortiz02} and SIMBAD.
\footnotemark[$X$] IRAM telescope pointing coordinates.
\end{tabnote}
\end{sidewaystable*}

\begin{figure*}
\begin{center}
\resizebox{0.99\hsize}{!}{\includegraphics[angle=0]{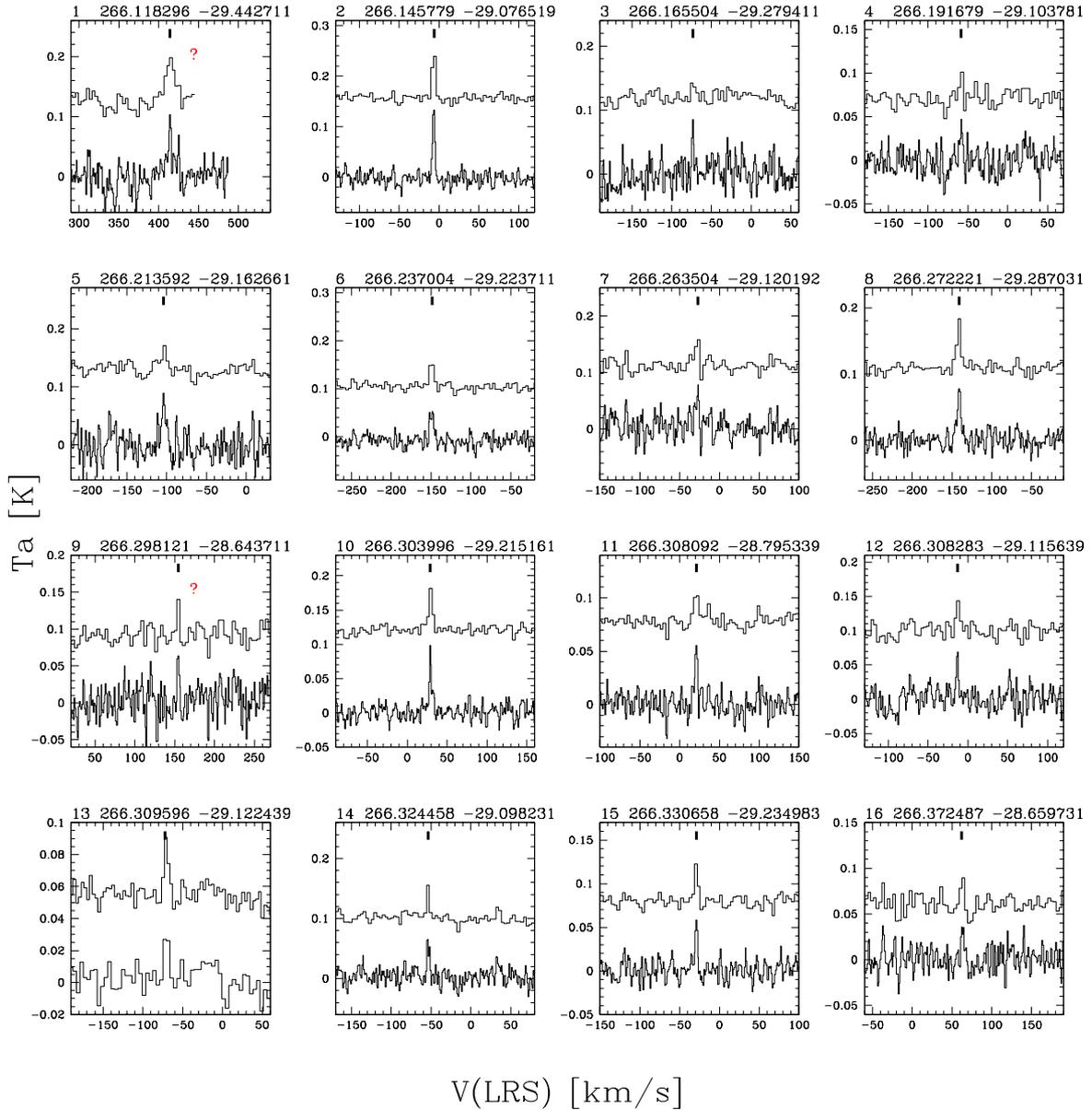}}
\end{center}
\caption{\label{chartes} IRAM spectra of the detected  lines. 
The IDs and observed coordinates (in J2000, degrees) are taken from Table
\ref{table:detections} and printed on top.
In each panel, the spectrum from the auto-correlator (1.1 \kms\ 
channel separation)
is plotted at the bottom, while the spectrum from the filterbank (3.5
\kms\ channel separation)
is plotted at the top. The small black marker at the top shows the
derived stellar velocity \emph{using the SiO transition rest frequency.}
} 
\end{figure*}

\addtocounter{figure}{-1}
\begin{figure*}
\begin{center}
\resizebox{0.99\hsize}{!}{\includegraphics[angle=0]{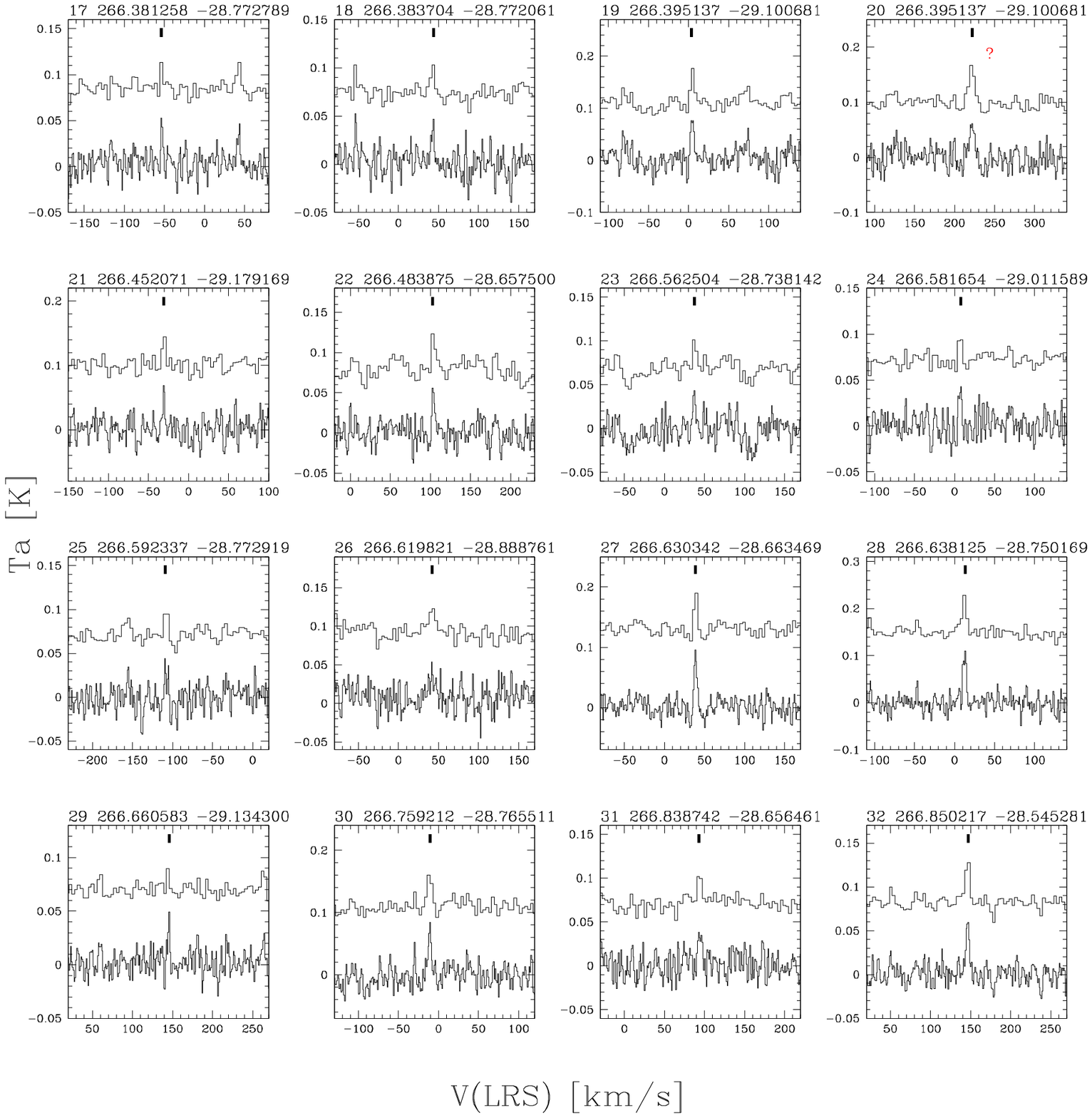}}
\end{center}
\caption{ Continuation of Figure \ref{chartes}.} 
\end{figure*}

\begin{figure*}
\begin{centering}
\resizebox{0.33\hsize}{!}{\includegraphics[angle=0]{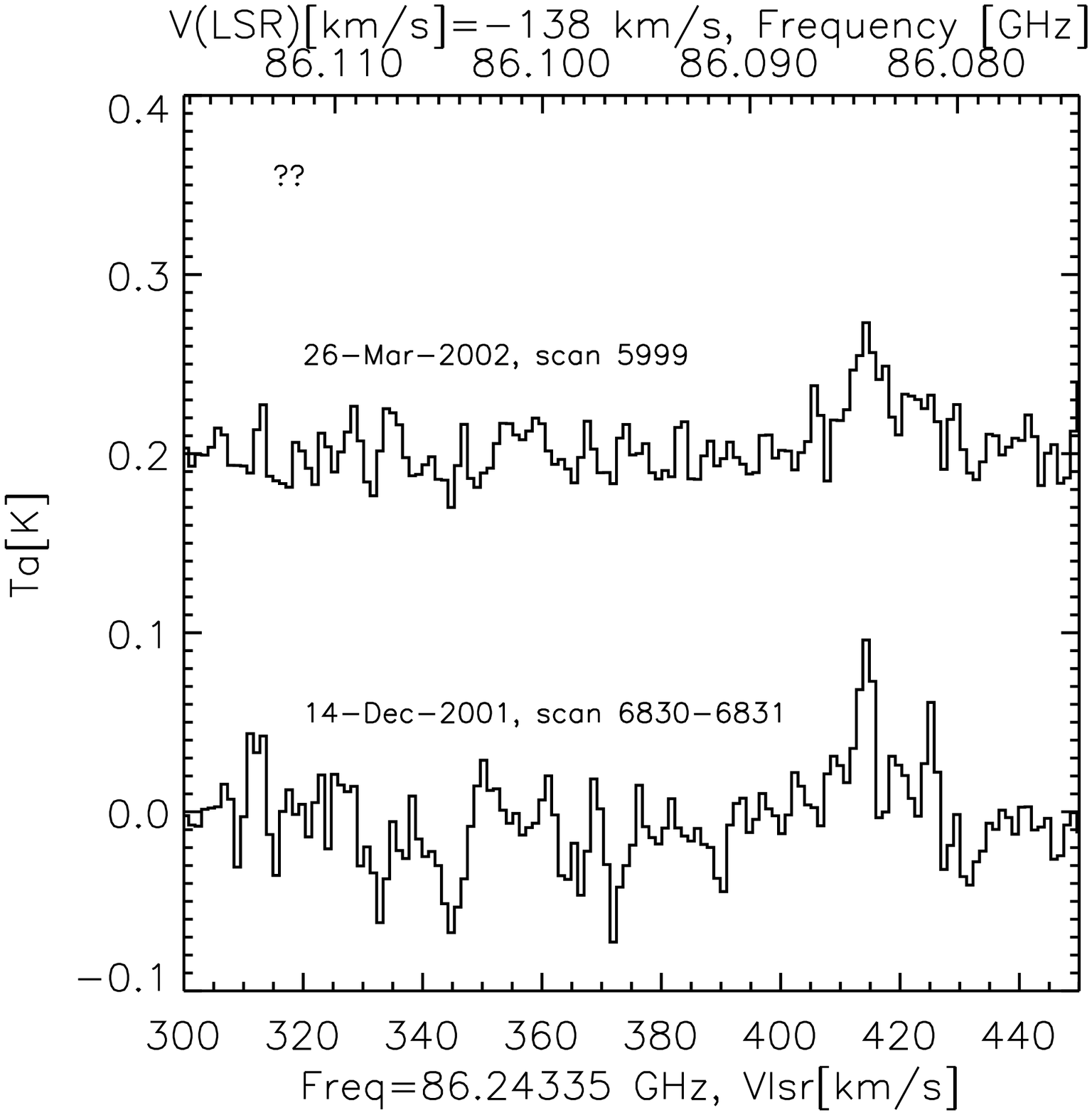}}
\resizebox{0.33\hsize}{!}{\includegraphics[angle=0]{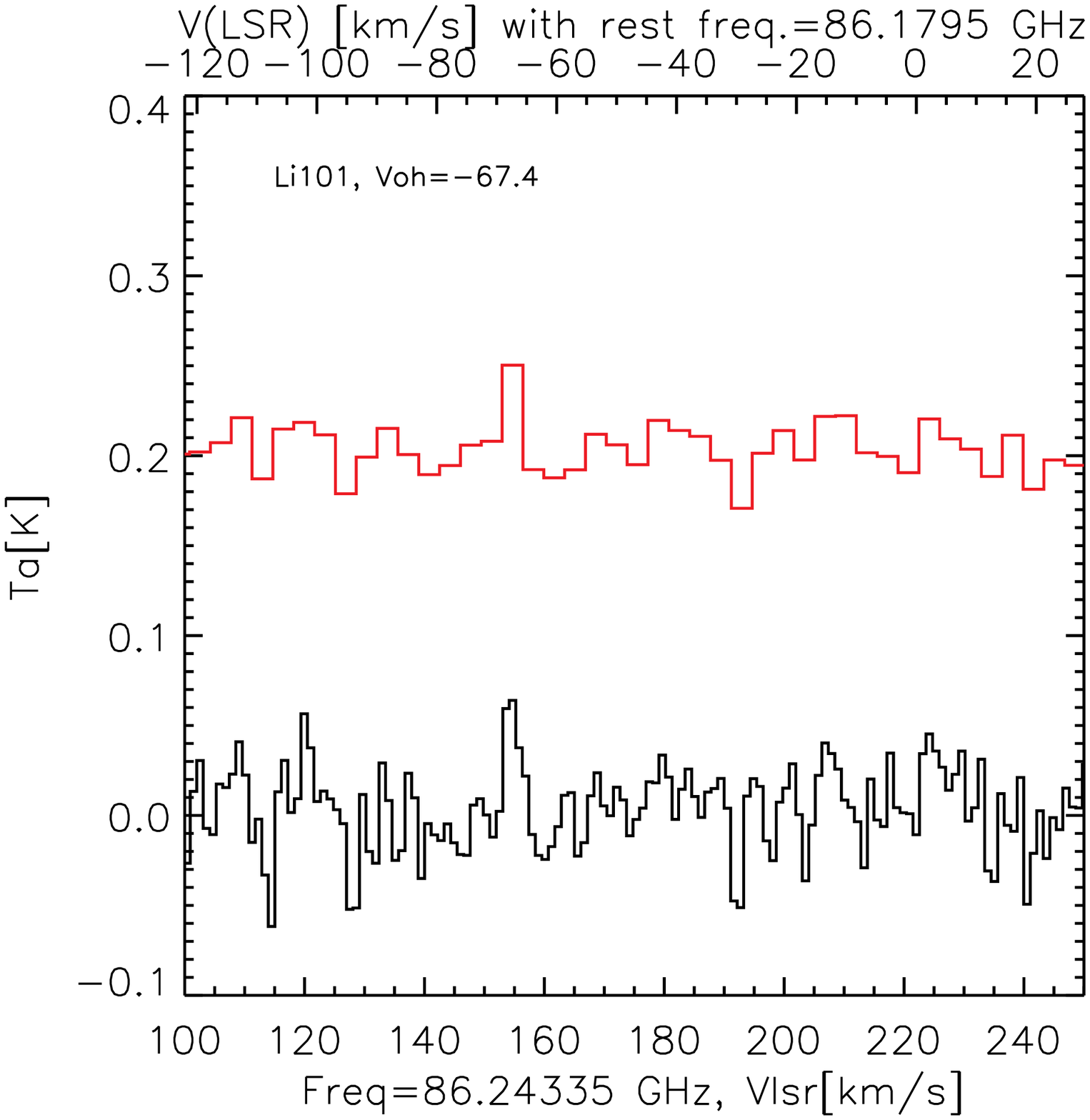}}
\resizebox{0.33\hsize}{!}{\includegraphics[angle=0]{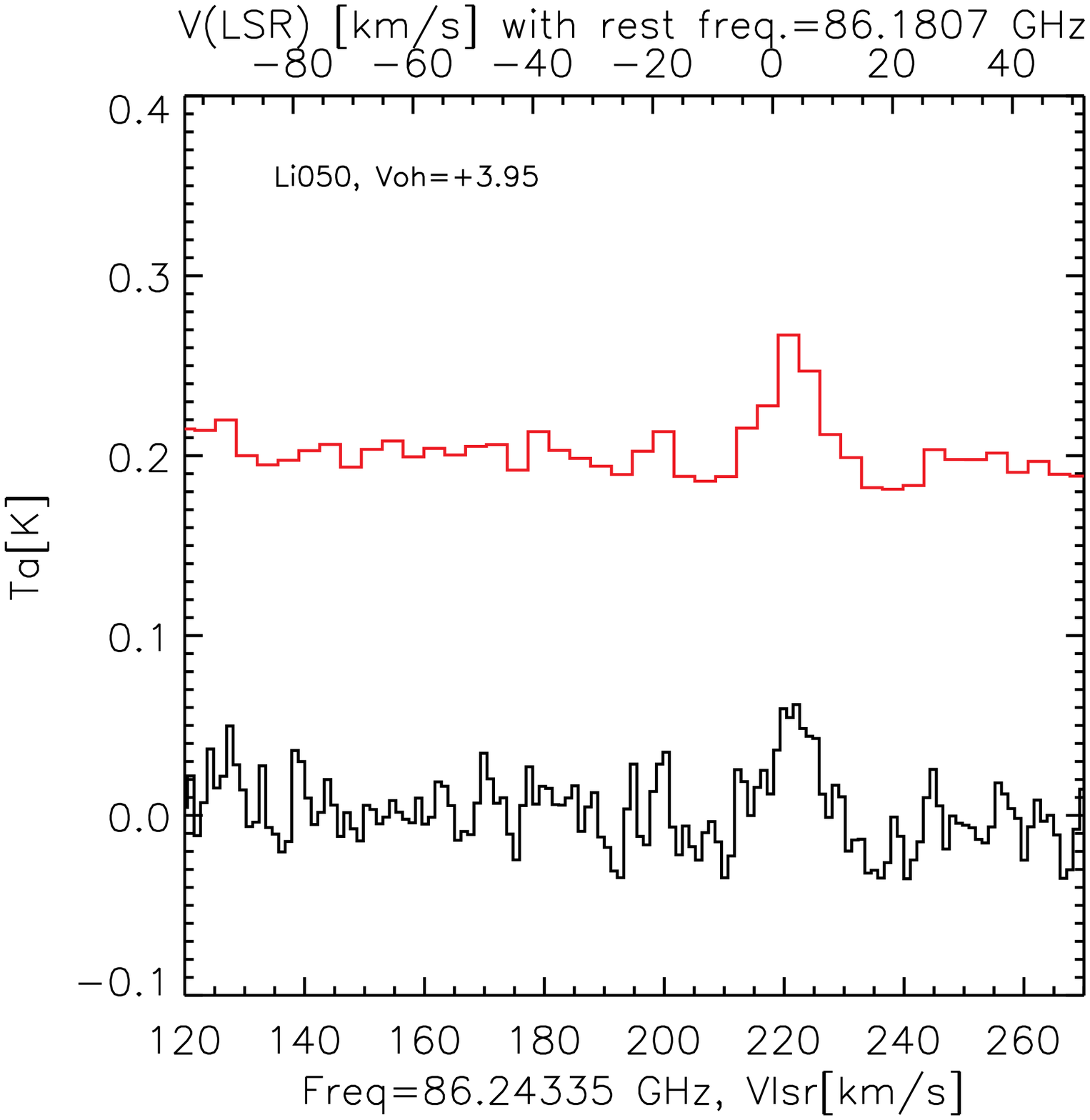}}
\end{centering}
\caption{\label{newline} 
{\it Left panel:} Spectra of the high-velocity line in the beam of Li004.
Two spectra   obtained with the auto-correlator in two different epochs 
are plotted in black. By assuming it arises from the OH/IR star
(OH \Vlsr=$-138.2$ \kms\ at 1612 MHz), the line frequency  is estimated 
on the top axis.
{\it Middle and right panels:} Spectra of a second detection in the  beams of 
Li101 and Li050. 
By assuming the  SiO rest frequency, the new lines are shifted by  
$\Delta V_{\rm LSR}=220\pm2.8$ \kms\ with respect 
to the OH lines  at 1612 MHz. The exact frequencies obtained by assuming 
the velocities of the OH/IR stars are given in the figure titles.
When using the rest frequency of the CCS line (86.181 GHz) we obtain
Delta V(LSR) = 217 \kms, and with the OH molecular transition at 86.178 GHz
 $\Delta V_{\rm LSR}=227$ \kms.
 }
\end{figure*}

\section{Detected lines} 
\label{analysis}
 In order to measure the emission from the maser  lines, 
 a linear baseline was subtracted from the spectra.
  This was a detection project and the typical  signal-to-noise ratio
  threshold  of detection for the
  expected velocity resolution is over 2.5-3 $\sigma$ over multiple
  channels near the centroid of the OH emission in
  the (1.1 \kms) autocorrelator spectra.
For each line, parameters for e.g.\ peak and FWHM  were estimated using a  simple 
Gaussian fit and listed in Tables \ref{table:detections} and 
\ref{table:non-detections}. 

We searched for 86 GHz SiO maser emission in 67 GC OH/IR stars and
detected  32 lines as presented in Table  \ref{table:detections} 
and Fig.\ \ref{chartes}. 
A total of 28 OH/IR stars were detected
in the SiO $(v=1, J = 2 \rightarrow 1)$  transition at 86.24335 GHz. 

\begin{figure*}
\begin{center}
\resizebox{0.8\hsize}{!}{\includegraphics[angle=0]{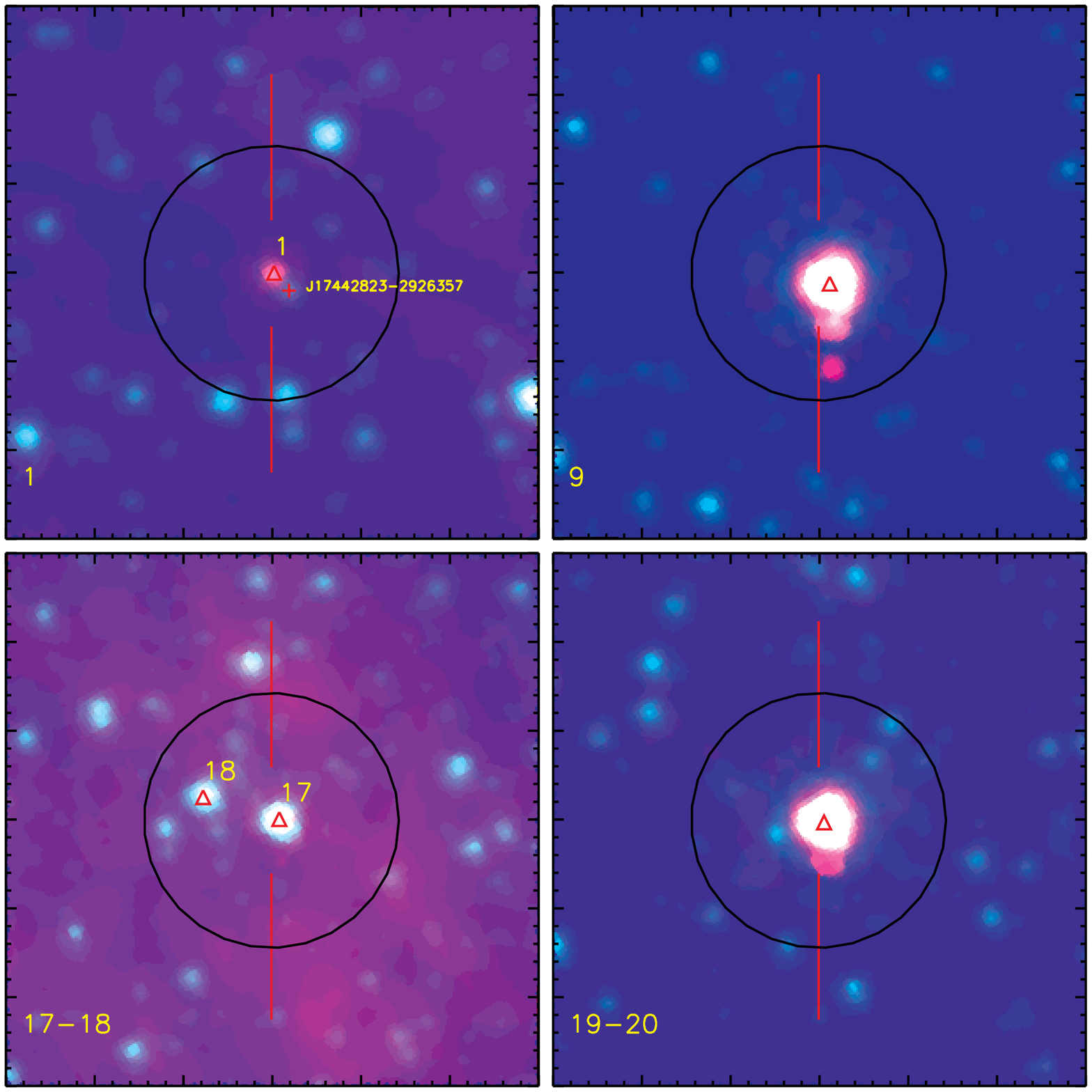}}
\end{center}
\caption{ \label{chances} Composite image. In red the GLIMPSE 8 \um, in green the GLIMPSE 4.5 \um, and
 in blue the UKIDSS K-band image. Identification numbers are as in Tables
1 and 2 of the paper. North is up and East to the left. 
The large black circle is centred on the IRAM pointing and has a diameter of 29\arcsec. 
The red triangles  mark the targeted OH/IR stars and the V4502 Sgr (ID=18)
variable, clockwise from the upper left respectively source ID 1,
source ID 9, source IDs 19/20 and source IDs 17/18. 
The red cross mark the location of 2MASS J17442823-2926357, which is near Li004 (ID=1) 
(not detected in $K$ band).
} 
\end{figure*}

SiO maser emission   was serendipitously  detected from the Mira-type star V4502Sgr 
(ID=18 in Table \ref{table:detections}), which appeared in the beam of ID=\#17, LI091.
For three positions (ID=1,9,20; alias LI004, LI101, LI050), possible 86 GHz lines are 
detected at velocities not coincident with those expected for the OH/IR stars 
(Fig.\ \ref{newline}).
In these cases, we visually inspected the ISOGAL and GLIMPSE  catalogs,
but no other bright  stars were seen at 8 \um\ (Fig.\ \ref{chances}).
As the offsets from the expected SiO maser velocities 
of LI101 and LI050 are very similar (i.e., 222 and 218 \kms)
we suggest that these indicate  other line detections  from the OH/IR stars themself at a 
frequency of about 86.18 GHz. 
A line at 86.181 GHz (a CCS-molecule transition, $J_N=6_7 - 5_6$) has been detected towards the 
carbon AGB star IRC+10216 that is also an OH masing star
\citep{cernicharo87,yamamoto90}. However, as 
one of these OH/IR stars  has an SiO maser, the association with a
carbon-rich star is unlikely  \citep[see also][]{stroh18}. Since, recently,
detections of highly excited OH main-line transitions in AGB stars have been
reported by \citet{khouri19}, we suggest another highly excited  OH transition
(e.g.. J$=\frac{17}{2}$ ($v=1$) F=8$^+$-8$^-$ at 86.178 GHz\footnote{ 
 See the online catalog of molecular lines at $https://spec.jpl.nasa.gov/.$}). \\
At the location of LI004 (\Vlsr=$-138.20$ \kms) a broad  line (FWHM$\approx11$ \kms) 
was detected at \Vlsr=+414.3 \kms\ when assuming SiO emission  (seen in both epochs).
Detections of other transitions are needed to confirm the nature of this line 
that could arise from  the envelope of the highest velocity  AGB star
measured in the GC.

\section{Available photometric magnitudes} 

OH/IR stars are among the  brightest   mid-infrared sources; indeed,
a large number of OH/IR stars were already seen with the IRAS mid-infrared satellite.
They are easily  detected even in the  crowded Galactic centre  region \citep{ortiz02}.
The observed OH/IR stars are located within ISOGAL fields
and coincide  with bright 15 \um\ point sources \citep{omont03,schuller03,ortiz02}.
\citet{wood98} was the first to photometrically monitor at near-infrared wavelength 
the Galactic centre OH/IR stars and to discover that they mostly coincided with 
long-period variables.   \citet{wood98}  provided near-infrared magnitudes of the  identified
long-period variables (CASPIR pixel scale was 0\farcs25). 
They were found within 2\arcsec\ from the radio positions of the OH masers.
While the mid-infrared measurements of OH/IR stars are unambiguously retrieved, 
by being the brightest sources of the field, near-infrared counterparts 
are often too faint to be detected.

In Table \ref{table.magnitudes}, we revised the mid-infrared measurements 
of the stars  by searching within The Midcourse Space Experiment (MSX) 
\citep[ with a sensitivity in $A$-band of 0.1 Jy and a 
maximum flux measured of  $\approx700$ Jy,][]{egan03,price01}, 
The Wide-field Infrared Survey Explorer (WISE) \citep[ with a sensitivity of $\approx0.9$ mJy (3.8 mag) and  
a  saturation at $\approx 140$ Jy (-3.0 mag) at 11.6 \um,][ and The WISE All-Sky online catalog]{wright10},  
MIPS Galactic Plane Survey (MIPSGAL) \citep{gutermuth15}, and 
Galactic Legacy Infrared Midplane Survey Extraordinaire (GLIMPSE) catalogs 
\citep[ with  fluxes from  $\approx$10 mJy 
to 1.6 Jy at at $8$ \um\ in v2,][]{churchwell09,benjamin03}.
Pointing positions were  taken from a preliminary
version of the ISOGAL catalog \citep[within 1\arcsec,][]{omont03}, and are
listed in Tables \ref{table:detections} and \ref{table:non-detections}.
For all pointing positions  a unique MSX counterpart was found within a search radius of 
14\farcs5 (the IRAM  half beam size). MSX has a spatial resolution of 18\farcs3 in $A$-band; 
the average distances from our positions and the MSX centroids is of 3\farcs0 with a 
$\sigma$ of 2\farcs2. MSX point sources were not available in correspondence 
of four OH maser positions. 
Flux at 8 \um\ (A-band) ranges from $\sim$ 0.15 to $\sim$ 11.0 Jy,
with a mean of 1.6 Jy ($\sigma$=1.4 Jy).\\
WISE imaged the sky with a spatial resolution of 6\arcsec\ and an astrometric 
accuracy  within 0\farcs4 \citep{wright10}. 
By  using a search radius to 10\arcsec,  we found 64  WISE matches to our 67 OH/IR stars.
We searched for the closest GLIMPSE counterparts by  adopting a search radius of 10\arcsec.
To avoid false identifications, knowing that  OH/IR stars are 
among the brightest and reddest mid-infrared point sources \citep{ortiz02},
we first considered data points with [8.0] $<$ 7.0 mag.
A total of 58 GLIMPSE valid matches to the 67 OH/IR stars were retrieved, 
plus a match for V4502 Sgr.

Average near-infrared magnitudes were provided by \citet{wood98},  
\citet{glass01},  and \citet{matsunaga09} for 57 targets and periods for 53 targets 
(see Tables \ref{table:detections}, \ref{table:non-detections}, and \ref{table.magnitudes}).

As stated above, near-infrared counterparts have historically been found
by detecting long-period variables, but  coordinates with a subarcsec accuracy are often missing. 
Nowadays, with the availability of accurate 
astrometry from several near-infrared Galactic plane surveys, it is possible to retrieve them
by positional coincidence.
We retrieved the original near-infrared 2MASS coordinates 
\citep[$JHK_s$,][]{twomass} 
of the matches provided by WISE and GLIMPSE.
In highly-crowded regions at the resolution of GLIMPSE 
near-infrared misidentification are possible, especially for the faintest stars.
We checked the 2MASS positions and GLIMPSE positions
with the source centroids of the GLIMPSE 8 \um\ and UKIDSS K-band
chartes. UKIDSS  images have a spatial resolution of 0\farcs2 \citep{lucas08}.
Only for stars \#33, \#46, \#55, \#66, \#68, GLIMPSE coordinates had to 
be astrometrically improved, 
by taking the UKIDSS data point coinciding with the GLIMPSE centroid
in order  to avoid misidentification.

With the improved coordinates,   we searched in the  General Catalogue 
of Variable Stars (GCVS) \citep{gcvs}. 47 of the 67 OH/IR stars were identified within 
a search radius of 1\arcsec, plus two other secure identifications at a larger separation
(see Table \ref{table.magnitudes} footnotes).
One of the near-infrared counterparts identified by 
\citet{wood98} appeared incorrect
(see the details on star \#1 in the  Sect.\ \ref{serend}).

For most of these stars the stellar spectral energy distribution, 
based on average near-infrared magnitudes plus several mid-infrared bands,
was analysed by \citet{wood98} and \citet{ortiz02} 
and estimates of extinction and bolometric fluxes 
are provided in their works. 
Furthermore, the near-infrared spectra taken by  \citet[][]{vanhollebeke06} 
confirm strong water vapour  absorption, which is the typical signature of
large amplitudes AGB variables.

\subsection{Serendipitous 86 GHz detections: ID=\#1, \#9, \#18, and \#20} 
\label{serend}

For  positions with two 86 GHz lines detected (Li091 and Li050) 
and with 86 GHz lines at significantly different velocities than that of the known OH masers 
(Li004 and  Li101), we carefully inspected the GLIMPSE and UKIDSS images.  
Maser emitters are usually associated with bright mid-infrared data points
(see Fig.\ \ref{chances}).

At the position of the  OH/IR star Li004 (ID=\#1), which is not seen at 86 GHz, 
a line is detected at 415 \kms. Li004 is the brightest mid-infrared star  with  
[4.5]=10.4 mag, [5.8]=7.8 mag, and [8.0]=6.11 mag, [15.0]=4.24 mag. At this location 
no  UKIDSS data point is seen, the GLIMPSE images show a  nearby  peak at a small 
separation of $\approx 2\farcs5$ with [3.6] =9.969 mag, [4.5]=9.730 mag, 
[5.8]=9.444 mag	(not detected at 8 \um). 
The ISOGAL/DENIS  (\Ks=11.5 mag) position  coincides with this other source seen 
at 3.6 \um\  (which is not centred on the  8 \um\ source) and with the 
2MASS J17442823-2926357 (\Ks=11.5 mag),  UKIDSS K-band 
data point (UGPSJ174428.23-292635.7, K=11.43 mag). 
\citet{wood98} did not detect any variability and reported 
an average $K$=11.48 mag, i.e they analysed 2MASS J17442823-2926357, a false
counterpart to the OH/IR stars.
 Only two other 8 \um\ GLIMPSE stars fall within the IRAM 
beam. They are faint  with [8.0]=7.70 and 7.80 mag (\Ks = 9.75 and 10.61 mag).

With a 8\farcs8 separation between Li091 (ID=17) and V4502Sgr (ID=18), 
 both stars fall within the IRAM beam. 
Li091, which is the OH/IR star, coincides  
with the bright 15 \um\ star detected by ISOGAL \citep[3.6 mag,][]{ortiz02}. 
Both variables are detected by GLIMPSE with [8.0]=5.47 and 6.65 mag, and 
[3.6-8.0]= 1.66 and 0.63 mag, respectively.

A second 86 GHz line is detected at the position of Li050 (lines ID=\#19 and \#20), and
a line at an unexpected velocity is found in the direction of Li101 (ID=\#9).
There are no other bright mid-infrared stars ([8.0]$<$ 8.9 mag) in these positions.
The sources were not observed by \citet{stroh19} (Stroh private communication, 2020).
There are no other reports of these  line detections towards  these sources.

\begin{sidewaystable*} 
\vspace*{+0.5cm} 
\caption{\label{table.magnitudes}   Infrared measurements of the targeted OH/IR stars 
and of the serendipitous detection of  V4502Sgr. } 
\begin{center}
{\tiny
\begin{tabular}{@{\extracolsep{-.09in}} l|rrlr|rrr|rr|rr|rrrr|rrr|r|rrrrrrrrrrrrrrrrrrrrr}
\hline 
 & \multicolumn{4}{c}{\rm NIR positions} &    \multicolumn{3}{c}{\rm Averages}   &
 \multicolumn{2}{c}{\rm GCVS}&\multicolumn{2}{c}{\rm ISOGAL}& \multicolumn{4}{c}{\rm GLIMPSE}&
 \multicolumn{4}{c}{\rm MSX}& 
  \multicolumn{4}{c}{\rm WISE}  & \multicolumn{1}{c}{\rm MIPSGAL}     \\ 
\hline 
 {\rm ID} & RA & DEC    &  Survey &   2MASS \Ks & $<J>$ & $<H>$ & $<K>$ & Sep & GCVS & [7] &[15]&{\rm [3.6]} & {\rm [4.5]} & {\rm [5.8]} & {\rm [8.0]} &
 {\it A}  & {\it C}  &{\it D}  &{\it E}  &
 {\it W1} &{\it W2}  & {\it W3} &  {\it W4} & {\it 24\um} \\ 
\hline 
          &    &      &          &{\rm [mag]} & {\rm [mag]} &{\rm [mag]}&{\rm [mag]} &\arcsec& &	 {\rm [mag]} &  \arcsec &{\rm [mag]} & {\rm [mag]}     & {\rm [mag]} &{\rm [mag]}  &  {\rm [mag]} &{\rm [mag]} 
 &{\rm [mag]}  & {\rm [mag]}&{\rm [mag]}&{\rm [mag]}&{\rm [mag]}&{\rm [mag]}&{\rm [mag]}\\
\hline

                             1  &  17 44 28.37& $-29  $  26 33.71  &GLIMPSE  &$..$   &$..$  & 13.60?  & 11.48?  & $..$   &      none  & 6.53  &  4.24  & $..$  & 10.37  &  7.80  &  6.11  &  5.71  & 98.98  &  3.92  & 98.98  & $..$  & $..$  & $..$  & $..$  &  2.42  \\
                             2  &  17 44 34.96& $-29  $  04 35.67  &  2MASS  & 9.24   &$..$  & $..$  & $..$  &   0.0   & V4458~Sgr  & 1.30  & -0.48  & 88.88  & 88.88  & 88.88  & 88.88  &  1.83  &  0.52  & -0.05  & -0.83  &  6.07  &  3.19  &  1.58  & -0.29  & $..$  \\
                             3  &  17 44 39.67& $-29  $  16 46.38  &  2MASS  &11.03   &$..$  & $..$  & 10.87  &   0.1   & V4462~Sgr  & 4.50  &  2.35  &  7.16  &  6.05  &  4.47  &  4.13  &  3.94  &  2.85  &  2.26  &  1.67  &  7.63  &  5.01  &  3.25  &  1.47  &  1.90  \\
                             4  &  17 44 45.96& $-29  $  06 13.63  &  2MASS  &11.22   &$..$  & 14.34  & 10.66  &   0.0   & V4465~Sgr  & 3.17  &  2.37  &  6.69  &  4.81  &  3.73  &  3.04  &  3.83  &  2.71  &  1.95  &  1.36  &  7.09  &  4.92  &  3.04  &  1.12  &  1.07  \\
                             5  &  17 44 51.25& $-29  $  09 45.31  &  2MASS  &11.21   &$..$  & $..$  & 11.46  &   0.0   & V4466~Sgr  & 4.44  &  2.85  &  7.76  &  6.39  &  5.28  &  4.69  &  4.30  &  3.26  &  2.74  &  2.29  &  8.54  &  6.43  &  4.70  &  1.65  &  2.03  \\
                             6  &  17 44 56.83& $-29  $  13 25.51  &  2MASS  & 7.77   &$..$  & $..$  & $..$  &   0.0   & V4470~Sgr  & 3.14  &  2.06  &  5.33  &  4.61  &  3.93  &  3.24  &  3.54  &  2.33  &  1.78  &  1.03  &  5.92  &  3.68  &  2.47  &  0.28  &  1.41  \\
                             7  &  17 45 03.20& $-29  $  07 12.64  &  2MASS  &10.42   &$..$  & 12.85  &  9.71  &   0.0   & V4473~Sgr  & 4.83  &  3.53  &  6.61  & $..$  &  4.27  &  4.18  & $..$  & $..$  & $..$  & $..$  &  4.41  &  4.84  &  3.29  &  1.62  &  2.29  \\
                             8  &  17 45 05.29& $-29  $  17 13.42  &  2MASS  & 6.88   &12.03  &  9.11  &  7.41  &   0.1   & V4475~Sgr  & 4.21  &  2.73  & 88.88  & 88.88  & 88.88  & 88.88  &  3.91  &  2.69  &  2.23  &  0.38  &  6.34  &  5.52  &  3.20  & -0.05  & $..$  \\
                             9  &  17 45 11.45& $-28  $  38 38.53  &  2MASS  &11.27   &$..$  & $..$  & 11.85  &   0.1   & V4481~Sgr  & 3.44  &  1.41  &  6.69  &  4.71  &  3.73  &  3.01  &  3.33  &  2.12  &  1.46  &  0.84  &  7.34  &  4.93  &  3.08  &  1.05  &  1.19  \\
                            10  &  17 45 12.92& $-29  $  12 54.67  &  2MASS  & 9.67   &$..$  & 12.98  &  9.68  &   0.1   & V4482~Sgr  & 3.49  &  2.66  & $..$  & $..$  & $..$  & $..$  &  3.61  &  2.32  &  1.96  &  1.48  &  6.30  &  4.27  &  2.73  &  1.24  &  1.09  \\
                            11  &  17 45 13.91& $-28  $  47 43.20  &  2MASS  &10.57   &$..$  & 13.85  & 10.34  &   0.1   & V4489~Sgr  & 3.43  &  1.43  & $..$  & $..$  & $..$  & $..$  &  3.59  &  2.40  &  1.82  &  1.27  &  6.54  &  4.28  &  2.49  &  0.66  &  1.01  \\
                            12  &  17 45 13.88& $-29  $  06 55.40  &GLIMPSE  &$..$   &$..$  & $..$  & 13.48  & $..$   &      none  & 3.35  &  2.23  &  9.55  &  7.18  &  5.32  &  4.32  &  3.40  &  2.41  &  1.56  &  1.23  &  9.03  &  6.54  &  3.48  &  1.44  & $..$  \\
                            13  &  17 45 14.27& $-29  $  07 20.75  &  2MASS  &11.13   &$..$  & 14.54  & 11.14  &   0.1   & V4487~Sgr  & 4.14  &  2.40  & $..$  & $..$  & $..$  & $..$  &  4.45  &  3.11  &  2.65  &  2.84  &  7.76  &  6.00  &  4.19  &  2.66  &  2.33  \\
                            14  &  17 45 17.83& $-29  $  05 53.34  &  2MASS  &10.57   &$..$  & 14.25  & 10.46  &   0.0   & V4492~Sgr  & 4.19  &  2.91  &  7.77  &  6.58  &  5.39  &  4.50  &  4.17  &  3.12  &  2.43  &  1.68  &  8.16  &  5.69  &  3.25  &  0.64  &  1.07  \\
                            15  &  17 45 19.34& $-29  $  14 05.79  &  2MASS  & 8.48   &$..$  & 12.25  &  8.97  &   0.0   & V4494~Sgr  & 4.08  &  2.02  &  5.27  &  4.71  &  3.88  &  3.31  &  3.60  &  2.44  &  2.03  &  1.43  &  6.05  &  4.14  &  2.52  &  0.74  &  1.78  \\
                            16  &  17 45 29.42& $-28  $  39 34.96  &  2MASS  &10.80   &$..$  & 13.51  & 10.63  &   0.1   & V4499~Sgr  & 4.55  &  2.50  &  6.88  &  6.12  &  4.48  &  4.00  &  3.83  &  2.64  &  2.02  &  1.55  &  6.29  &  4.22  &  2.41  &  0.75  &  0.85  \\
                            17  &  17 45 31.43& $-28  $  46 21.89  &  2MASS  & 8.96   &14.46  & 10.88  & $..$  &   0.3   & V4501~Sgr  & 4.87  &  3.60  &  7.13  &  6.49  &  5.87  &  5.47  & 97.97  & 97.97  & 97.97  & 98.98  & $..$  & $..$  & $..$  & $..$  & $..$  \\
                            18  &  17 45 32.09& $-28  $  46 19.42  &  2MASS  & 8.57   &$..$  & $..$  & $..$  &   0.0   & V4502~Sgr  &$..$  & $..$  &  7.28  &  7.10  &  6.61  &  6.65  & 97.97  & 97.97  & 97.97  & 98.98  & $..$  & $..$  & $..$  & $..$  & $..$  \\
                            19  &  17 45 34.79& $-29  $  06 02.68  &  2MASS  & 8.98   &15.91  & 11.47  &  9.05  &   0.1   & V4507~Sgr  & 3.56  &  2.55  &  5.59  &  4.75  &  3.90  &  3.25  &  3.90  &  2.69  &  2.18  &  2.08  &  6.06  &  4.31  &  2.87  &  1.03  & $..$  \\
                            20  &  17 45 34.79& $-29  $  06 02.68  &  2MASS  & 8.98   &15.91  & 11.47  &  9.05  &   0.1   & V4507~Sgr  & 3.56  &  2.55  &  5.59  &  4.75  &  3.90  &  3.25  &  3.90  &  2.69  &  2.18  &  2.08  &  6.06  &  4.31  &  2.87  &  1.03  & $..$  \\
                            21  &  17 45 48.48& $-29  $  10 45.20  &  2MASS  & 9.74   &$..$  & 12.58  &  9.57  &   0.0   & V4519~Sgr  & 4.90  &  3.52  &  6.63  &  5.82  &  4.34  &  4.11  &  4.61  &  3.40  &  2.96  &  2.52  &  7.63  &  5.72  &  3.93  &  2.27  &  2.01  \\
                            22  &  17 45 56.08& $-28  $  39 27.46  &  2MASS  & 9.79   &$..$  & $..$  & 10.48  &   0.1   & V4532~Sgr  &$..$  &  1.88  &  6.84  &  6.16  &  4.63  &  4.20  &  3.40  &  2.08  &  1.72  &  1.19  &  5.87  &  3.93  &  2.49  &  0.58  &  1.02  \\
                            23  &  17 46 14.96& $-28  $  44 17.34  &  2MASS  & 9.24   &15.35  & 11.02  &  8.76  &   0.0   & V4549~Sgr  & 4.56  &  3.07  &  6.77  &  6.28  &  5.03  &  4.72  &  4.63  &  3.27  &  2.49  &  0.81  &  7.39  &  5.98  &  3.81  &  0.31  & $..$  \\
                            24  &  17 46 19.51& $-29  $  00 41.34  &  2MASS  &10.63   &$..$  & $..$  & 11.68  & $..$   &      none  & 4.89  &  2.58  &  7.72  &  6.34  &  5.11  &  4.41  &  5.20  &  3.89  &  2.75  &  1.54  &  7.08  &  4.77  &  3.09  &  0.72  & $..$  \\
                            25  &  17 46 22.11& $-28  $  46 22.61  &  2MASS  & 9.19   &$..$  & 13.14  &  9.93  &   0.0   & V4551~Sgr  & 3.84  &  1.97  &  6.84  &  5.99  &  4.40  &  4.07  &  3.77  &  2.55  &  2.00  &  0.86  &  7.18  &  5.76  &  4.61  & -2.13  & $..$  \\
                            26  &  17 46 28.71& $-28  $  53 19.49  &  2MASS  & 8.15   &15.67  & 11.48  &  8.85  &   0.0   & V4555~Sgr  & 4.54  &  2.64  &  6.67  &  5.56  &  4.62  &  4.17  &  4.64  &  3.18  &  2.87  &  2.76  &  6.48  &  5.07  &  3.48  &  1.61  &  1.45  \\
                            27  &  17 46 31.23& $-28  $  39 48.68  &  2MASS  & 8.26   &$..$  & 11.62  &  8.84  & $..$   &      none  & 4.86  &  3.09  &  6.68  &  6.20  &  5.05  &  4.71  &  4.12  &  2.67  &  2.35  &  0.81  &  7.33  &  6.23  &  5.09  & 98.98  &  2.01  \\
                            28  &  17 46 33.07& $-28  $  45 01.04  &  2MASS  &10.08   &$..$  & $..$  & 10.49  & $..$   &      none  & 4.89  &  4.04  &  7.12  &  6.18  &  4.86  &  4.55  & $..$  & $..$  & $..$  & $..$  & $..$  & $..$  & $..$  & $..$  &  2.94  \\
                            29  &  17 46 38.51& $-29  $  08 01.22  &GLIMPSE  &95.95   &$..$  & $..$  & $..$  & $..$   &      none  & 4.66  &  2.64  &  7.27  &  6.10  &  5.04  &  4.42  &  4.56  &  3.27  &  2.80  &  2.62  &  6.66  &  4.93  &  3.25  &  1.40  &  2.00  \\
                            30  &  17 47 02.17& $-28  $  45 55.79  &  2MASS  & 8.78   &$..$  & 13.10  &  9.94  &   0.0   & V4570~Sgr  & 4.32  &  2.88  &  7.04  &  6.28  &  5.08  &  4.50  &  5.14  &  3.69  &  3.09  & 98.98  &  7.34  &  5.76  &  4.12  &  1.88  &  1.20  \\
                            31  &  17 47 21.28& $-28  $  39 22.84  &  2MASS  &10.74   &$..$  & $..$  & 11.52  &   0.1   & V4573~Sgr  & 3.64  &  1.88  &  7.63  &  6.30  &  5.09  &  4.45  &  3.59  &  2.36  &  1.79  &  1.28  &  6.67  &  4.59  &  2.96  &  0.97  &  0.95  \\
                            32  &  17 47 24.02& $-28  $  32 42.93  &  2MASS  & 8.14   &$..$  & 11.99  &  8.80  &   0.1   & V4574~Sgr  & 3.48  &  2.12  &  5.62  &  4.55  &  3.82  &  3.24  &  3.48  &  2.29  &  1.84  &  1.42  &  7.16  &  5.39  &  3.58  &  1.49  &  1.90  \\
\end{tabular}
}
\begin{list}{}{}
\item[]{\bf Magic values:} \\
$..$ not available.\\ 
88.88 retrieved match false, star not extracted, image saturation.\\ 
95.95 2MASS (provided by WISE or GLIMPSE) or ISOGAL-DENIS data points 
      were false near-infrared matches and removed. \\ 
97.97 Confusion. \\ 
98.98 upper limit.\\
\end{list}
\end{center}
\end{sidewaystable*} 

\addtocounter{table}{-1}
\begin{sidewaystable*} 
\vspace*{+0.5cm} 
\caption{Continuation of Table \ref{table.magnitudes}.    } 
\begin{center}
{\tiny
\begin{tabular}{@{\extracolsep{-.09in}} l|rrlr|rrr|rr|rr|rrrr|rrr|r|rrrrrrrrrrrrrrrrrrrrr}
\hline 
 & \multicolumn{4}{c}{\rm NIR positions} &    \multicolumn{3}{c}{\rm Averages}   &
 \multicolumn{2}{c}{\rm GCVS}&\multicolumn{2}{c}{\rm ISOGAL}& \multicolumn{4}{c}{\rm GLIMPSE}&
 \multicolumn{4}{c}{\rm MSX}& 
  \multicolumn{4}{c}{\rm WISE}  & \multicolumn{1}{c}{\rm MIPSGAL}     \\ 
\hline 
 {\rm ID} & RA & DEC    &  Survey &   2MASS \Ks & $<J>$ & $<H>$ & $<K>$ & Sep & GCVS & [7] &[15]&{\rm [3.6]} & {\rm [4.5]} & {\rm [5.8]} & {\rm [8.0]} &
 {\it A}  & {\it C}  &{\it D}  &{\it E}  &
 {\it W1} &{\it W2}  & {\it W3} &  {\it W4} & {\it 24\um} \\ 
\hline 
          &    &      &          &{\rm [mag]} & {\rm [mag]} &{\rm [mag]}&{\rm [mag]} &\arcsec& &	 {\rm [mag]} &  \arcsec &{\rm [mag]} & {\rm [mag]}     & {\rm [mag]} &{\rm [mag]}  &  {\rm [mag]} &{\rm [mag]} 
 &{\rm [mag]}  & {\rm [mag]}&{\rm [mag]}&{\rm [mag]}&{\rm [mag]}&{\rm [mag]}&{\rm [mag]}\\
\hline 
                            33  &  17 43 45.48& $-29  $  26 17.22  & UKIDSS  &95.95   &$..$  & $..$  & $..$  & $..$   &      none  & 4.06  &  2.14  & $..$  &  7.24  &  5.03  &  4.09  &  4.47  &  3.04  &  1.97  &  1.44  &  9.52  &  6.44  &  2.77  &  0.66  & $..$  \\
                            34  &  17 43 53.95& $-29  $  25 23.43  &GLIMPSE  &95.95   &$..$  & $..$  & $..$  & $..$   &      none  & 4.48  &  1.80  &  7.53  &  6.07  &  3.94  &  3.09  &  4.37  &  3.38  &  2.25  &  1.49  &  8.19  &  4.92  &  2.47  &  0.27  &  1.08  \\
                            35  &  17 44 14.95& $-28  $  45 06.02  &  2MASS  & 8.49   &13.86  & 10.31  &  8.13  &   0.1   & V4449~Sgr  & 4.24  &  1.95  &  6.70  &  6.17  &  4.77  &  4.10  &  3.80  &  2.43  &  2.15  &  1.16  &  6.82  &  5.77  &  3.99  &  1.94  &  1.89  \\
                            36  &  17 44 34.48& $-29  $  10 37.23  &GLIMPSE  &95.95   &$..$  & $..$  & 11.21  & $..$   &      none  & 4.05  &  1.76  &  7.20  &  5.16  &  4.67  & $..$  &  3.95  &  2.84  &  2.03  &  1.45  &  6.95  &  5.07  &  3.06  &  0.77  &  1.02  \\
                            37  &  17 44 44.43& $-29  $  05 37.92  &  2MASS  & 9.92   &$..$  & 14.00  & 10.40  &   0.0   & V4463~Sgr  & 3.57  &  2.05  &  7.44  &  6.32  &  5.18  &  4.56  &  3.77  &  2.61  &  1.91  &  1.39  &  7.19  &  5.00  &  2.98  &  1.08  &  1.10  \\
                            38  &  17 44 47.94& $-29  $  06 49.93  &  2MASS  & 9.39   &14.92  & 11.28  &  9.43  & $..$   &      none  & 5.03  &  2.85  &  7.81  &  7.20  &  5.73  &  4.71  &  4.81  &  3.62  &  2.55  &  2.23  &  7.64  &  6.45  &  4.09  &  1.90  &  1.48  \\
                            39  &  17 44 54.16& $-29  $  13 44.88  &  2MASS  & 8.52   &15.53  & 11.07  &  8.67  &   0.0   & V4468~Sgr  & 4.89  &  3.23  & 88.88  & 88.88  & 88.88  & 88.88  &  3.78  &  2.28  &  1.71  & -0.13  &  6.89  &  6.08  &  4.09  &  1.65  & $..$  \\
                            40  &  17 44 57.74& $-29  $  20 42.42  &  2MASS  &11.91   &$..$  & $..$  & 10.85  &   0.1   & V4471~Sgr  & 4.66  &  2.45  &  6.75  &  5.92  &  4.12  &  3.46  &  4.34  &  3.03  &  2.31  &  1.06  &  8.66  &  6.36  &  4.05  &  0.30  & $..$  \\
                            41  &  17 45 01.70& $-29  $  02 49.90  &  2MASS  & 8.54   &13.47  & 10.26  &  8.28  &   0.0   & V4779~Sgr  & 4.91  &  3.26  &  6.69  &  6.13  &  5.10  &  4.57  &  4.22  &  2.89  &  2.70  &  2.26  &  6.35  &  5.08  &  3.56  &  1.65  &  2.21  \\
                            42  &  17 45 06.90& $-29  $  03 33.40  &GLIMPSE  &95.95   &14.64  & 11.11  &  9.04  & $..$   &      none  & 4.86  &  2.73  &  7.15  &  6.45  &  5.77  &  5.22  &  5.08  &  3.24  &  2.39  &  1.02  &  6.52  &  5.09  &  3.24  &  0.34  & $..$  \\
                            43  &  17 45 10.39& $-29  $  18 12.26  &  2MASS  &12.65   &$..$  & 14.80  & 11.11  &   0.1   & V4480~Sgr  & 3.98  &  1.80  &  7.84  &  6.76  &  4.92  &  4.23  &  4.38  &  3.34  &  2.49  &  1.59  &  7.28  &  4.48  &  2.67  &  0.43  & $..$  \\
                            44  &  17 45 12.45& $-28  $  40 44.39  &  2MASS  & 9.80   &$..$  & 11.47  &  8.69  &   0.0   & V4484~Sgr  & 4.68  &  1.36  &  7.72  &  6.40  &  5.16  &  4.18  &  3.57  &  2.24  &  1.78  &  0.97  &  8.73  &  6.84  &  3.71  &  1.50  &  1.37  \\
                            45  &  17 45 13.07& $-29  $  09 36.33  &  2MASS  & 9.72   &15.35  & 11.51  &  9.38  &   0.0   & V4483~Sgr  & 4.29  &  2.83  &  6.82  &  5.70  &  4.99  &  4.46  &  4.34  &  3.27  &  2.72  &  2.06  &  6.44  &  5.04  &  4.44  &  3.17  &  2.46  \\
                            46  &  17 45 13.86& $-29  $  15 28.73  & UKIDSS  &95.95   &$..$  & $..$  & 11.25  &   2.1   & V4485~Sgr$^a$  & 3.47  &  1.59  & $..$  & $..$  & $..$  & $..$  &  3.24  &  2.11  &  1.50  &  0.33  &  6.65  &  3.97  &  2.52  & -0.68  & $..$  \\
                            47  &  17 45 16.43& $-29  $  15 37.61  &  2MASS  & 7.66   &14.26  & 10.54  &  8.42  &   0.0   & V4490~Sgr  & 4.60  &  3.48  &  6.68  &  6.13  &  5.06  &  4.74  &  3.79  &  2.61  &  2.44  &  1.00  &  6.33  &  5.18  &  3.27  &  0.63  & $..$  \\
                            48  &  17 45 18.04& $-29  $  18 03.91  &GLIMPSE  &95.95   &$..$  & $..$  & 13.67  &   0.5   & V4491~Sgr  & 5.91  &  4.32  &  8.67  &  6.46  &  5.41  &  4.99  &  6.61  & 98.98  &  4.73  & 98.98  & 10.47  &  6.54  &  5.61  &  3.14  &  3.50  \\
                            49  &  17 45 29.24& $-29  $  07 04.25  &  2MASS  & 9.54   &$..$  & 12.70  & 10.09  &   0.1   & V4875~Sgr  & 4.40  &  2.96  &  7.32  &  6.55  &  5.65  &  5.20  &  4.67  &  3.33  &  2.96  &  2.13  &  6.84  &  5.79  &  4.19  &  1.64  &  1.96  \\
                            50  &  17 45 29.46& $-29  $  09 16.42  &GLIMPSE  &$..$   &$..$  & $..$  & 11.41  & $..$   &      none  & 4.87  &  3.06  &  7.82  &  6.67  &  5.49  &  4.95  &  4.78  &  3.90  &  3.22  &  2.57  &  8.46  &  6.27  &  4.41  &  1.61  &  1.44  \\
                            51  &  17 45 33.26& $-29  $  11 23.20  &GLIMPSE  &$..$   &$..$  & $..$  & 12.54  & $..$   &      none  & 5.34  &  3.81  &  9.75  &  7.58  &  6.22  &  5.54  & $..$  & $..$  & $..$  & $..$  & 98.98  &  6.81  &  4.87  &  2.53  &  3.24  \\
                            52  &  17 45 33.44& $-28  $  43 45.00  &  2MASS  &10.83   &$..$  & $..$  & 10.78  & $..$   &      none  & 4.00  &  2.07  &  6.88  &  5.82  &  4.26  &  4.05  &  3.79  &  2.52  &  2.18  &  1.54  &  6.67  &  5.17  &  2.64  &  0.68  &  1.83  \\
                            53  &  17 45 34.41& $-29  $  12 54.12  &  2MASS  & 9.93   &$..$  & $..$  & $..$  & $..$   &      none  & 4.30  &  2.71  &  6.85  &  6.38  &  4.87  &  4.43  &  5.30  &  3.79  &  3.08  &  2.42  &  6.23  &  4.49  &  3.09  &  1.30  &  2.06  \\
                            54  &  17 45 46.57& $-28  $  32 39.50  &  2MASS  &10.35   &$..$  & 13.43  &  9.36  &   0.0   & V4518~Sgr  & 3.36  &  1.73  & $..$  & $..$  & $..$  & $..$  &  3.64  &  2.50  &  2.00  &  1.37  &  7.32  &  5.32  &  3.47  &  1.29  &  1.33  \\
                            55  &  17 45 54.21& $-28  $  31 47.06  & UKIDSS  &11.49   &$..$  & $..$  &  9.62  &   1.1   & V4529~Sgr$^b$  & 3.85  &  1.66  & 88.88  & 88.88  & 88.88  & 88.88  &  3.25  &  2.04  &  1.25  &  0.92  &  7.48  &  5.11  &  2.97  &  0.84  & $..$  \\
                            56  &  17 46 00.92& $-29  $  01 23.24  &GLIMPSE  &$..$   &$..$  & $..$  & 14.79  &   0.3   & V4538~Sgr  & 2.30  &  0.59  &  7.00  &  4.59  &  2.63  & $..$  &  3.75  &  2.50  &  1.63  &  1.08  &  9.04  &  5.44  &  2.17  &  0.11  & $..$  \\
                            57  &  17 46 11.68& $-28  $  59 32.44  &  2MASS  & 9.40   &$..$  & 13.20  &  9.58  &   0.0   & V5020~Sgr  & 5.23  &  3.87  &  7.16  &  6.37  &  5.42  &  4.89  &  4.93  &  3.92  &  3.12  &  2.22  &  7.56  &  5.87  &  3.72  &  0.37  &  2.53  \\
                            58  &  17 46 14.48& $-28  $  36 39.54  &  2MASS  &12.02   &$..$  & $..$  & 12.62  &   0.1   & V4548~Sgr  & 4.28  &  3.29  &  8.81  &  7.16  &  5.90  &  5.24  &  4.40  &  3.38  &  2.90  &  2.53  &  7.94  &  5.72  &  4.18  &  2.13  &  2.29  \\
                            59  &  17 46 15.25& $-28  $  55 42.36  &GLIMPSE  &$..$   &$..$  & $..$  & $..$  &   0.1   & V4550~Sgr  & 3.24  &  1.55  &  9.20  &  6.98  &  5.31  &  4.42  &  3.67  &  2.52  &  1.65  &  1.16  &  8.55  &  5.49  &  3.01  &  0.14  & $..$  \\
                            60  &  17 46 15.74& $-28  $  56 32.27  &  2MASS  &10.09   &$..$  & 14.03  & 10.05  &   0.0   & V5039~Sgr  & 5.30  &  3.31  &  7.15  &  6.04  &  4.87  &  4.33  &  4.99  &  3.61  &  2.68  &  1.46  &  7.81  &  5.93  &  3.97  &  0.64  & $..$  \\
                            61  &  17 46 24.87& $-29  $  00 01.41  &  2MASS  &10.97   &$..$  & 15.22  & 11.21  &   0.0   & V4552~Sgr  & 4.72  &  2.80  &  7.70  &  6.40  &  5.27  &  4.70  &  4.62  &  3.40  &  2.59  &  2.02  &  7.00  &  4.84  &  3.42  &  0.63  &  1.25  \\
                            62  &  17 46 30.71& $-28  $  31 32.81  &GLIMPSE  &95.95   &$..$  & 12.47  &  9.65  & $..$   &      none  & 7.03  &  5.77  &  8.22  &  7.13  &  6.79  &  6.70  & $..$  & $..$  & $..$  & $..$  &  8.25  &  7.20  &  6.37  &  3.48  &  4.16  \\
                            63  &  17 46 31.68& $-28  $  35 41.05  &GLIMPSE  &95.95   &$..$  & 14.83  & 11.60  &   0.3   & V4559~Sgr  & 4.56  &  3.11  &  7.90  &  6.39  &  5.35  &  4.86  &  4.84  &  3.73  &  3.28  & 98.98  &  6.47  &  5.97  &  5.19  &  1.73  &  2.33  \\
                            64  &  17 46 35.31& $-28  $  58 57.05  &GLIMPSE  &$..$   &$..$  & $..$  & $..$  &   0.2   & V4562~Sgr  & 4.26  &  2.26  &  8.16  &  6.18  &  4.13  &  3.11  &  3.43  &  2.24  &  1.50  &  0.74  &  9.10  &  6.47  &  2.90  &  0.31  &  1.45  \\
                            65  &  17 46 42.29& $-28  $  33 26.13  &  2MASS  &11.41   &$..$  & 14.97  & 11.18  &   0.0   & V4567~Sgr  & 4.74  &  3.98  &  7.91  &  6.39  &  5.42  &  5.03  &  5.15  & 98.98  &  4.14  & 98.98  &  8.66  &  6.73  &  5.00  &  1.68  &  3.42  \\
                            66  &  17 46 44.84& $-28  $  34 59.66  & UKIDSS  &95.95   &$..$  & $..$  & $..$  & $..$   &      none  & 4.70  &  2.49  &  8.08  &  6.35  &  4.89  &  4.22  &  4.54  &  3.34  &  2.87  &  2.87  &  7.83  &  5.09  &  3.21  &  1.38  &  1.20  \\
                            67  &  17 46 47.81& $-28  $  47 15.16  &GLIMPSE  &$..$   &$..$  & $..$  & 13.62  &   0.3   & V4568~Sgr  & 5.72  &  3.50  &  9.90  &  8.14  &  6.59  &  5.55  &  5.49  &  3.95  &  3.42  &  3.07  & 10.60  &  7.95  &  4.52  &  1.25  &  2.61  \\
                            68  &  17 46 58.95& $-28  $  17 00.56  & UKIDSS  &95.95   &$..$  & $..$  & $..$  & $..$   &      none  & 4.06  &  1.67  &  8.37  &  6.27  &  4.77  &  4.03  &  3.65  &  2.29  &  1.75  &  0.99  &  8.27  &  5.14  &  2.87  &  0.66  & $..$  \\
                            69  &  17 47 39.65& $-28  $  35 46.78  &GLIMPSE  &$..$   &$..$  & $..$  & 13.47  &   0.8   & V4575~Sgr  &$..$  &  2.20  &  6.71  &  4.76  &  3.65  & $..$  &  2.86  &  1.86  &  1.13  &  0.54  &  6.95  &  3.69  &  1.98  &  0.18  &  0.99  \\

\end{tabular}
}
\end{center}
\begin{tabnote}
$^a$ Star \#46  coincides (0\farcs05) with  V4485 Sgr (Hmag=14.299 mag, Ksmag=10.714 mag
and $\Delta$K=2.030 mag) which is analyzed in 
the VVV catalog of Galactic Bulge Type II Cepheids NIR data
by \citet{braga19} (note that the GCVS has the old  coordinates, 2\farcs1 away).  
$^b$ The 2MASS centroid of star \#55 was  recentered.
This star coincide with V4529 Sgr (note that the GCVS has the old coordinate, 1\farcs1 away).
\end{tabnote}

\end{sidewaystable*}

\section{SiO masers and properties of the OH/IR stars}

A global detection rate of 42\% is obtained for 86 GHz SiO masers in OH/IR stars.

\subsection{SiO maser detections and MSX colors: difference between Miras and OH/IR stars.}

\begin{figure}
\begin{center}
\resizebox{0.48\hsize}{!}{\includegraphics[angle=0]{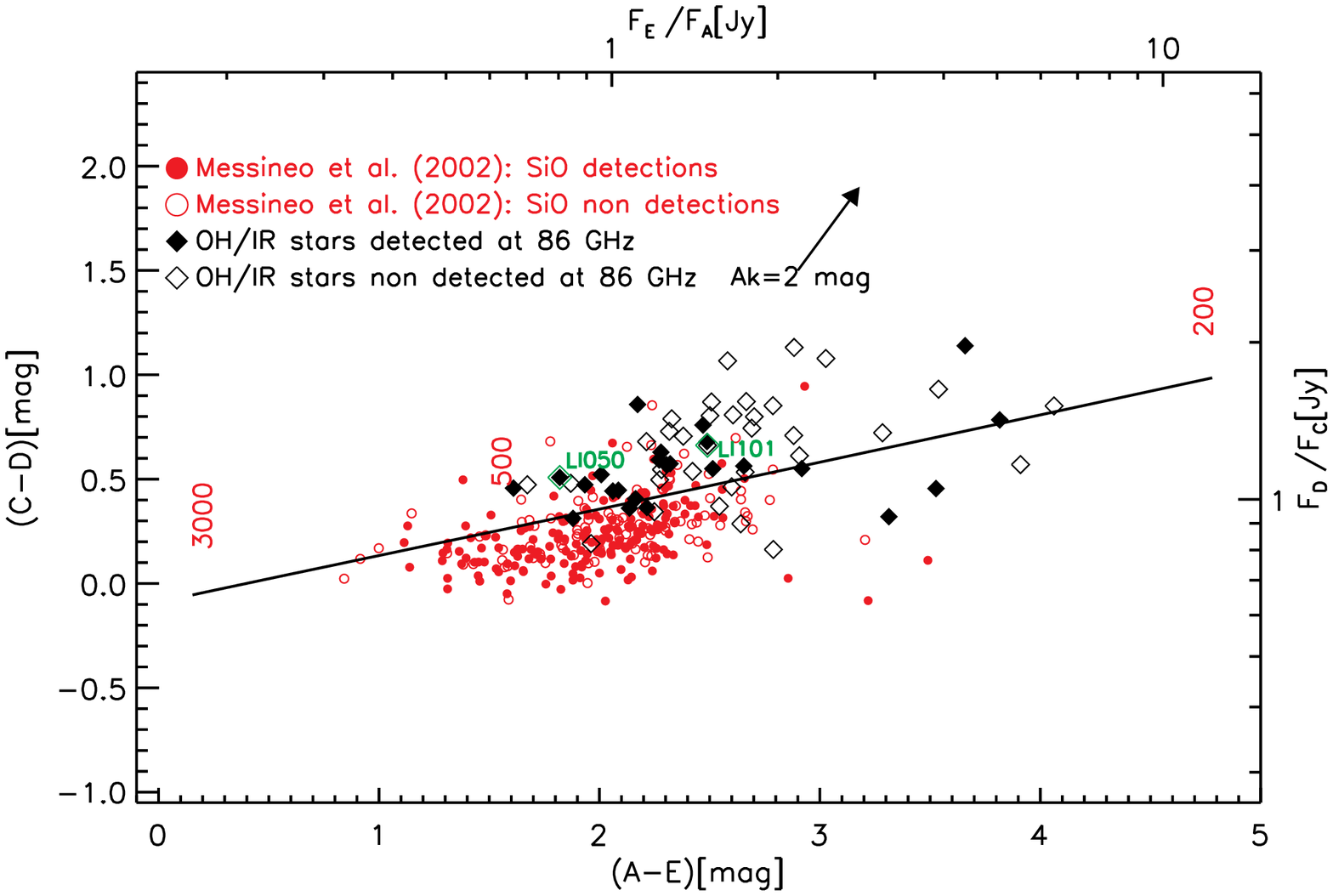}}
\resizebox{0.48\hsize}{!}{\includegraphics[angle=0]{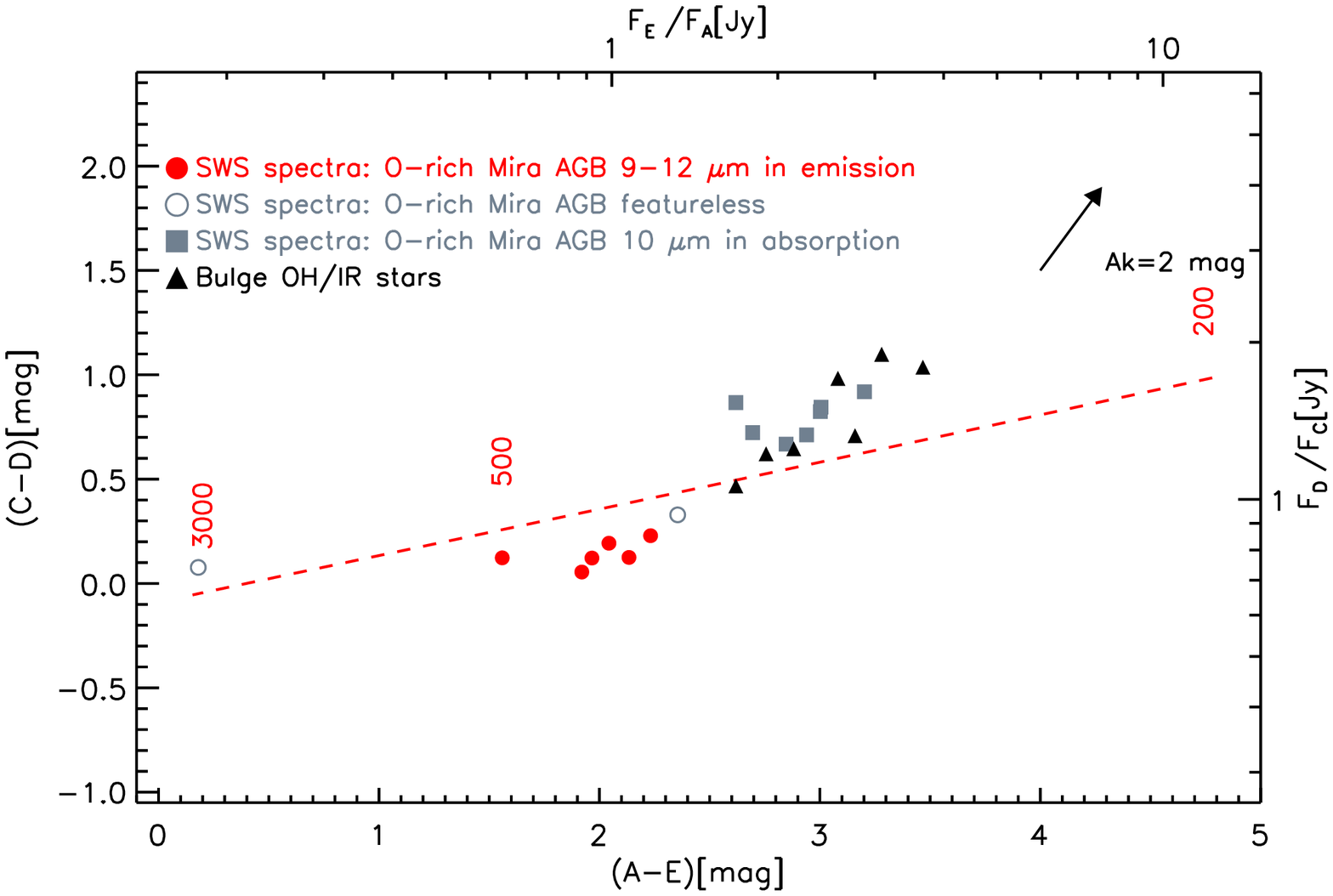}}
\end{center}
\caption{\label{msxcol} {\bf Left panel:} 
MSX  ($C-D$) versus ($A-E$) colors. 
Black diamonds mark the OH/IR stars analysed in this work (the locations of Li101 and Li050
are marked in green with two green labels),
while red circles mark the Mira-like stars of \citet{messineo02}. 86 GHz SiO maser detections
are marked with filled symbols, non-detections with open symbols.
{\bf Right panel:} MSX $C-D$ versus $A-E$ plot of O-rich Mira stars with ISO-SWS spectra.
Red-filled circles mark
those ISO-SWS spectra with 10 \um\ silicate features in emission, while gray-filled squares
indicate those with the
silicate feature  in absorption. For comparison, the 7 Bulge 
OH/IR stars modeled in \citet{blommaert18} (with silicate features in absorption)
are plotted with  black-filled triangles.
} 
\end{figure}

OH/IR stars  are the most obscured long-period variables, with optically thick 
envelopes, as already inferred from studies of the mid-infrared colors 
from IRAS \citep[]{habing96}.

In our 2000-2003 search for 86 GHz maser emission (Messineo M., PhD thesis, 2004 University of Leiden),
we deliberately selected Mira-like stars and excluded OH/IR-like 
stars by targeting bluer colors \citep{messineo02}.
There were several reasons to do so; efficiency in increasing the 
number of stellar velocities in a manner complementary to OH surveys, and 
previous knowledge 
(at that time provided by, e.g., \citet{haikala94} and \citet{bujarrabal94})
that 86 GHz maser emission was more frequently 
detected in Mira stars than in thicker-shelled OH/IR stars.
With the presented dataset we can verify our  hypothesis, 
which were based on a very small number of sources, and possibly
even determine a change in the detection rate of SiO masers with stellar colors.

Mira-like stars have a color distribution indistinguishable from that
of the OH/IR stars   in the  $(D-E)$ versus $(A-C)$ diagram 
\citep[see Figures in][]{messineo18}. This diagram is useful only
to separate  mass-loss AGB stars from post-AGBs \citep{sevenster02}.
There is another combination of MSX filters that allows us to separate 
Mira-like AGB stars with thinner envelopes from the 
thicker-envelope OH/IR stars \citep{messineo04,lumsden02}.
In Fig.\ \ref{msxcol}, we show the  MSX $C-D$ versus $A-E$ 
colors\footnote{$C$, $D$, $A$, and $E$ indicated the MSX magnitudes.
The $A$-band covers from 6 to 11 \um, the $C$-band  from 11 to 13 \um, 
the $D$-band covers from 13.5 to 16 \um, the $E$-band covers from 
18 to 25 \um, as shown in Fig.\ 1 of \citet[][]{messineo05}.}.

The OH/IR stars  appear to have  redder $C-D$ colors and $A-E$ colors
than the comparison sample of Mira-like stars (without OH emission) 
from Messineo's thesis \citep{messineo04,messineo18}.
The colors are not corrected for interstellar extinction, however, 
for \Aks=3 mag corrections to the $C-D$ colors are within 0.3-0.6 mag and to the 
$A-E$ colors within 0.3-0.4 mag  \citep{messineo05}.
The bulk of Mira-like stars from \citet{messineo18}  \citep[see also][]{messineo02} 
have $A-E$ colors from 1.0 to 2.5 mag, while the OH/IR stars here studied
have redder $A-E$ colors from 1.5 to 4.0 mag.
For the Mira-like stars of \citet{messineo18}, 
the SiO detection rate is 66\% (261 targets and 172 detections) for colors $A-E < 2.5$ mag, 
but is equal to 43\% when considering the remaining Mira like stars (21 targets and 9 detections) 
with $2.5 < A-E < 3.5$ mag.
For the 29 OH/IR stars with $A-E < 2.5$ mag (average $C-D=0.54$ mag, $\sigma=0.16$ mag), 
the 86 GHz SiO detection rate is 59\%,  similar to that
of the Mira-like stars. For the 28 OH/IR stars with $A-E > 2.5$ mag 
(with average $C-D=0.70$ mag, $\sigma=0.26$ mag),
the 86 GHz detection rate is 25\%.
The 86 GHz SiO detection rate drops for sources redder than $A-E=2.5$ mag.
This is somewhat in line with the earlier findings of \citet{haikala94} and 
\citet{bujarrabal94} that 86 GHz maser emission occurs more frequently in 
Miras than in OH/IR stars. They must have looked at the reddest OH/IR stars.

The redder $C-D$ colors of the OH/IR stars 
are most likely due to the broad 10 \um\ silicate features (in absorption from $\approx 8$ to $\approx$ 12.5 \um) 
in combination with the MSX filter profiles \citep{messineo04}.
In Fig. \ref{msxcol}, we also plot the MSX colors of O-rich Miras and OH/IR stars with
available ISO-SWS spectra from the library of \citet{sloan03}.
O-rich Miras with 10 \um\ silicate spectra in emission have  $C-D$ colors located
below the black body curve, as expected by \citet{messineo04}. The $C-D$ colors 
of stars with 10 \um\ feature in absorption fall above the curve.
For comparison, the MSX colors of the Bulge OH/IR stars modeled by 
\citet{blommaert18} are shown, and  they also fall above the black body line.

Most of the sources here considered have $D-E < 1.38$ mag, 
which delimits the color region dominated by SiO masers 
described in \citet[with an SiO detection rate of 80\% at the high sensitivity of ALMA]{stroh19}.
Indeed, 99\% of the Mira-like stars in \citet{messineo18} 
have colours $ D-E < 1.38$ mag and
91\% of the OH/IR stars here studied 
(among the five redder OH/IR stars there are three  SiO maser detections).

The bluer Mira-like stars detected at 86 GHz have mass-loss rate from $10^{-7}$
to $2 \times 10^{-5}$ \Msun\ yr$^{-1}$ with a peak at $10^{-6}$-$10^{-5}$ 
\Msun\ yr$^{-1}$ \citep{messineo04}.
For Bulge OH/IR stars, \citet{ortiz02} estimate mass-loss rates 
from  $3 \times 10^{-6}$ to a few $10^{-5}$ \Msun\ yr$^{-1}$. 
The rare (seven) CO line detections in Bulge OH/IR stars yield 
estimates of mass-loss rates from 2 $\times 10^{-5}$  to  
9.5 $\times 10^{-5}$ \Msun\ yr$^{-1}$ \citep[][]{blommaert18}.

\subsection{Line ratios}

We searched the literature for previous detections of these SiO masers
at 43 GHz
\citep[e.g,][]{li10,fujii06,deguchi04,sjouwerman02,deguchi00a,deguchi00b,izumiura98,shiki97,lindqvist91}.
We found detections for 22 of the OH/IR stars only and non-detections
for four of the OH/IR stars.
There are 10 detections at both 86 and 43 GHz which yield
these average ratios $\frac{I(43 {\rm GHz,} v = 1)} {I(86 {\rm GHz,} v  = 1)}= 1.09$
and $\frac{I(43 {\rm GHz,} v = 2)} {I(86 {\rm GHz,} v = 1)}= 1.75$.
Despite the non simultaneity of the data, taken at a random 
phase, the mean is meaningful because   the photometric variations of the stars
are not synchronised one to another.
These values are similar to the quasi-simultaneous ratios analyzed by \citet{stroh18}
for a selection of thinner-shell Miras.

\subsection{ SiO detection rates and periods, \vexp, Mbol: difference between Miras and OH/IR stars.}
We analyzed the 86 GHz detections as a function of   the stellar periods, amplitudes
(see Tables \ref{table:detections} and \ref{table:non-detections}). 
Interestingly, it appears that 86 GHz SiO maser detections 
arise from OH/IR stars with periods longer than 500 days.
By restricting the analysis  to periods $> 500$ days, the 86 GHz
detection rate increases to 57\% (the global detection rate is 42\%). 
That the SiO maser detection rate is a steep function of periods 
had  already been concluded
 from the sample of Galactic centre large-amplitude
variables of \citet{glass01} and the 43 GHz SiO maser observations of \citet{imai02}.
Thereby, SiO maser emission traces  AGB variables with periods longer than 500 days.

All but one the 86 GHz SiO maser detections belong  to OH/IR stars with \vexp $> 14.5$ \kms.
The OH/IR stars  \vexp\ are known to increase with stellar periods \citep{lindqvist92}.

During their life, long-period variable stars lose mass at increasing rates,
thicken their envelopes, and lengthen their periods \citep{vassiliadis93}.
Periods depend on the stellar initial masses and 
ages of the variables.
The bulk of the Bulge OH/IR stars is made of  stars,
with  \Mbol\ from $-4.5$-$-5.0$ \Mbol\
\citep[][]{ortiz02,blommaert18}.
This \Mbol\ range suggests that most of the stars have initial masses between 1.5 and 2 \Msun. 
Because of this relatively narrow range in mass, the variety of observed 
stellar properties are dominated by evolution rather than initial masses 
\citep{ortiz02,blommaert18,qiao18}.
Bulge OH/IR stars do not follow the period-luminosity relation found for
Bulge Mira stars by \citet{glass95}, but they are systematically located below it.
The commonly accepted interpretation is that, during the pulsating phase, 
Miras  enter a regime of high-mass loss (superwind phase) where the AGBs
significantly stretch their stellar periods at an almost constant 
bolometric magnitude \citep[][]{vassiliadis93} departing from the locus 
of the Mira period-luminosity relation.
OH/IR stars have longer periods for a given luminosity than those found
in  Mira stars.
Of the sampled sources, 43 stars have
\Mbol\ estimates from the work of \citet{ortiz02} and \citet{wood98}.
They range from $-1.99$ to $-6.35$ mag, 20 detections and 23 non-detections. 
SiO maser detections occur towards stars brighter than  \Mbol $= -3.5$ mag.\\
Alternatively, \citet{urago20} explain these deviations 
from the period-luminosity relation 
at 3 \um\ to be attributed to circumstellar extinction. 
This strengthens the idea of \citet{sjouwerman09} that the optical thickness
is the key parameter to study OH/IR stars.

\begin{ack}
IRAM  is  supported   by  INSU/CNRS  (France),  MPG  (Germany) and   IGN   (Spain). 
MM carried part of this work  at the Leiden University in 2001-2002.
The work of MM from 2000 to 2004 was funded by the Netherlands
Research School for Astronomy (NOVA) through a netwerk 2, Ph.D. stipend. 
This work was partially supported by the National Natural Science Foundation of China 
(NSFC-  11421303, 11773025), and USTC grant KY2030000054.
The National Radio Astronomy Observatory is a facility of the National Science 
Foundation operated under cooperative agreement by Associated 
Universities, Inc. 

This publication makes use of data products from the Two Micron All Sky Survey, which 
is a joint project of the University of Massachusetts and the Infrared Processing and 
Analysis Center/California Institute of Technology, funded by the National Aeronautics 
and Space Administration and the National Science Foundation.
This work is based [in part] on observations made with the Spitzer Space Telescope, 
which is operated by the Jet Propulsion Laboratory, California Institute of Technology 
under a contract with NASA.
The DENIS project was supported, in France by the Institut National des Sciences de l'Univers, the Education Ministry 
and the Centre National de la Recherche Scientifique, in Germany by the State of Baden-Wuertemberg, in Spain by the 
DGICYT, in Italy by the Consiglio Nazionale delle Ricerche, in Austria by the Fonds zur Foerderung der 
wissenschaftlichen Forschung and the Bundesministerium fuer Wissenschaft und Forschung. 
This research made use of data products from the
Midcourse Space Experiment, the processing of which was funded by the Ballistic Missile Defence Organization with additional
support from the NASA office of Space Science. 
 This publication makes use of data products from
WISE, which is a joint project of the University of California, Los
Angeles, and the Jet Propulsion Laboratory/California Institute of Technology, 
funded by the National Aeronautics and Space Administration. 
This work is based on data obtained as part ofthe UKIRT Infrared Deep Sky Survey.
Based on observations with ISO, an ESA project with instruments funded by ESA Member 
States (especially the PI countries: France, Germany, the Netherlands and the United 
Kingdom) and with the participation of ISAS and NASA.
This research has made use of the SIMBAD data base, operated at CDS, Strasbourg,
France. This research made use of Montage, funded by the National Aeronautics and 
Space Administration’s Earth Science Technology Office, Computational Technnologies 
Project, under Cooperative Agreement Number NCC5-626 between NASA and the California 
Institute of Technology. The code is maintained by the NASA/IPAC Infrared Science Archive.
This research hasmade use of NASAs Astrophysics Data System BibliographicServices.
The author are grateful to the kind referee for his careful reading and constructive
suggestions. 
\end{ack}


\end{document}